\documentclass[aps,prd,twocolumn,fleqn,superscriptaddress,groupedaddress,nofootinbib,preprintnumbers]{revtex4}
\pdfoutput=1
\usepackage{amsmath,amssymb,graphicx,todonotes,array,booktabs}
\newcolumntype{L}[1]{>{\raggedright\let\newline\\\arraybackslash\hspace{0pt}}p{#1}}
\newcolumntype{C}[1]{>{\centering\let\newline\\\arraybackslash\hspace{0pt}}p{#1}}
\newcolumntype{R}[1]{>{\raggedleft\let\newline\\\arraybackslash\hspace{0pt}}p{#1}}

\usepackage[colorlinks = true,
            linkcolor = black,
            urlcolor  = blue,
            citecolor = black,
            anchorcolor = blue]{hyperref}

\begin{document}
\preprint{FERMILAB-PUB-19-192-T}
\preprint{LU-TP 19-14}
\preprint{MCNET-19-09}
\title{Simulation of vector boson plus many jet 
  final states at the high luminosity LHC}
\author{Stefan~H{\"o}che}
\affiliation{Fermi National Accelerator Laboratory,
  Batavia, IL, 60510-0500, USA}
\author{Stefan Prestel}
\affiliation{Department of Astronomy and Theoretical Physics,
  Lund University, S-223 62 Lund, Sweden}
\author{Holger Schulz}
\affiliation{Department of Physics,
  University of Cincinnati, Cincinnati, OH 45219, USA}

\begin{abstract}
  We present a novel event generation framework for the efficient 
  simulation of vector boson plus multi-jet backgrounds at the
  high-luminosity LHC and at possible future hadron colliders.
  MPI parallelization of parton-level and particle-level event
  generation and storage of parton-level event information using
  the HDF5 data format allow us to obtain leading-order merged
  Monte-Carlo predictions with up to nine jets in the final state.
  The parton-level event samples generated in this manner correspond
  to an integrated luminosity of 3${\rm ab}^{-1}$ and are made
  publicly available for future phenomenological studies. 
\end{abstract}

\maketitle

\section{Introduction}
The production of a di-lepton pair or a lepton-neutrino pair
at hadron colliders \cite{Drell:1970wh,Drell:1970yt} is both one of the most studied
and one of the best understood reactions in high-energy physics.
Calculations using fixed-order QCD perturbation theory have been performed
up to N$^2$LO accuracy \cite{Altarelli:1978id,
  Altarelli:1979ub,
  Hamberg:1990np,vanNeerven:1991gh}.
The production of an additional jet can be described fully exclusively at N$^2$LO accuracy
\cite{Ridder:2015dxa,Boughezal:2015ded,Boughezal:2016dtm,Ridder:2016nkl,Gehrmann-DeRidder:2017mvr},
and the production of up to five (four) additional jets in lepton-neutrino
(di-lepton) events can be described \cite{Giele:1993dj,Campbell:2002tg,Campbell:2003hd} at NLO accuracy,
partly in a fully automated fashion \cite{Ellis:2009zw,Berger:2009zg,Berger:2010vm,
  Berger:2010zx,Ita:2011wn,Bern:2013gka}.
Tree-level matrix element generators
\cite{Kanaki:2000ey,Papadopoulos:2000tt,Krauss:2001iv,Mangano:2002ea,
  Gleisberg:2008fv,Alwall:2011uj,Alwall:2014hca}
are capable of predicting the production rates of a Drell-Yan lepton pair
in association with any number of jets, limited only by computing power.
These predictions can be merged with parton showers \cite{
  Andre:1997vh,Mangano:2001xp,Catani:2001cc,Lonnblad:2001iq,
  Hamilton:2009ne,Hoeche:2010kg,Lonnblad:2012ng,Platzer:2012bs}
to create simulations that include the dominant effects of Sudakov resummation
and are at the same time leading order accurate at any jet multiplicity.
Such simulations prove to be an invaluable tool for LHC data analyses \cite{Buckley:2011ms}.

While techniques for the automated computation of tree-level matrix elements
have been devised long ago \cite{Berends:1987cv,Berends:1987me,Ballestrero:1994jn},
practical computations have not advanced considerably beyond the state of the art
of the early 2000s \cite{Mangano:2002ea,Gleisberg:2008fv}.
This has adverse effects on experimental measurements at the LHC,
where up to eight-jet final states are routinely probed~\cite{
  Aad:2013ysa,Aad:2014qxa,Aaboud:2017soa,Aaboud:2017hbk,
  Khachatryan:2016fue,Sirunyan:2017wgx,Sirunyan:2018cpw,Khachatryan:2016crw},
and nine-jet final states will be measured to excellent precision with an
integrated luminosity of $3{ab}^{-1}$.
We address this problem by providing a novel event generation framework,
combining core functionalities of Comix~\cite{Gleisberg:2008fv} and
Pythia~\cite{Sjostrand:2014zea}. Inspired by the preliminary studies
on HPC computing for perturbative QCD performed during the Snowmass 2013
community planning process \cite{Hoche:2013zja,Bauerdick:2014qka}
and by the successful parallelization of Alpgen~\cite{Childers:2015tyv},
we construct a workflow for parton-level and particle-level event generation
that provides a scalable solution from desktops to high-performance computers.
Existing limitations of Comix \cite{Benjamin:2017xdd} are addressed
and the new algorithms are tested in $W^\pm$ and $Z$ boson production with
up to nine jets at the LHC. The results of our parton-level event generation
campaign are publicly available~\cite{schulz_holger_2019_2678039,
  schulz_holger_2019_2678055,schulz_holger_2019_2678091}
and can be used as an input to particle-level simulations
in experimental analyses and phenomenology.
The code base can be downloaded from~\cite{hpcgenrepo}.

This paper is organized as follows. Section~\ref{sec:tech_intro}
introduces the technical challenges of particle-level simulations
at scale and Section~\ref{sec:framework} presents our new event generation framework.
Section~\ref{sec:results} discusses its computing performance and
presents some first physics results.
Section~\ref{sec:conclusions} contains an outlook.

\section{Efficient algorithms for high multiplicity multi-jet merging}
\label{sec:tech_intro}
Multi-jet merged event simulations at leading or next-to-leading order
QCD accuracy are the de facto standard for making precise, fully differential
signal and background predictions for Standard Model measurements and
new physics searches at the LHC~\cite{Buckley:2011ms}.
They provide a consistent combination of the fully differential resummation
provided by parton showers with exact higher-order perturbative QCD predictions
of events with resolved jets. Using the example of $pp\to Z$+jets, this implies
that the merging combines the parton-shower calculation of $pp\to Z$ with a
parton-shower calculation for $pp\to Z+j$, a
parton-shower calculation for $pp\to Z+jj$, and so on. As each of the results
to be merged is inclusive over additional QCD radiation, and the parton shower
can in principle populate the entire multi-jet phase space, two aspects
need to be addressed in any merging procedure:
\begin{enumerate}
\item The phase space of resolvable emissions in the parton shower must be restricted
  to the complement of the phase space in the fixed-order calculation. This procedure
  is called the {\it jet veto}, the observable used to separate the phase space
  is called the {\it jet criterion}, and the numerical value where the separation
  occurs is called the {\it merging scale}.
\item\label{step:meps_reweight_n_veto}
  The fixed-order result must be amended by the resummed higher-order
  corrections implemented by the parton shower, in order to maintain the
  logarithmic accuracy of the overall calculation. This involves
  \begin{enumerate}
  \item Re-interpreting the final-state configuration of the fixed-order calculation
    as having originated from a parton cascade~\cite{Andre:1997vh}. This procedure is called
    {\it clustering}, and the representations of the final-state configuration in terms of
    parton branchings are called {\it parton-shower histories}.
  \item Choosing appropriate scales for evaluating the strong coupling in each branching
    of this cascade, thereby resumming higher-order corrections to soft-gluon
    radiation~\cite{Amati:1980ch,Catani:1990rr}.
    This procedure is called {\it $\alpha_s$-reweighting}.
  \item Multiplying by appropriate Sudakov factors, representing the resummed unresolved
    real and virtual corrections~\cite{Catani:2001cc}. This is called
    {\it Sudakov reweighting}, and is usually implemented by
    {\it trial showers}~\cite{Lonnblad:2001iq}.
  \end{enumerate}
\end{enumerate}
Step~\ref{step:meps_reweight_n_veto} in this algorithm turns the inclusive
$pp\to Z+nj$ predictions into exclusive results, which describe the production
of {\it exactly} $n$ jets according to the jet criterion.\footnote{
  Note that the jet criterion need not correspond to an experimentally relevant
  jet algorithm. It may be a purely theoretical construct as long as the infrared
  limits are properly identified~\cite{Salam:2009jx}.}
They can then be added to obtain the merged result. Care has to be taken that
the result for the highest jet multiplicity remain inclusive over additional
radiation which is softer than the softest existing jet. This is known as the
{\it highest multiplicity treatment}.

\subsection{General Aspects of the Simulation}
\begin{figure}
  \includegraphics[width=\linewidth]{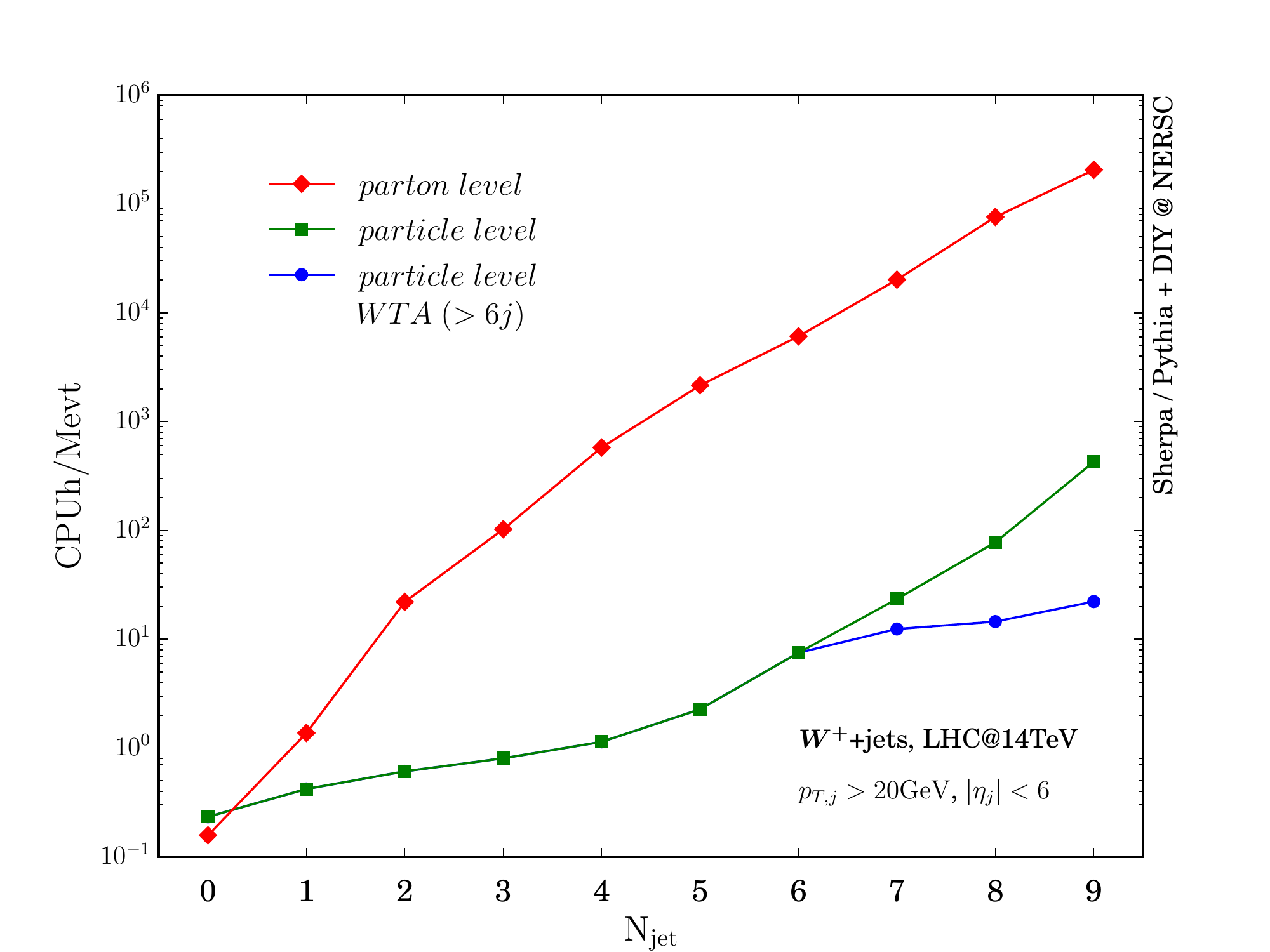}
  \caption{Scaling of computation time (in CPU hours per 1 million events)
    for parton-level and particle-level event generation in multi-jet merged
    computations of $W^++$jets at the LHC. We limit the number of quarks
    to $\le6$ in $W^++6,7$jet and to $\le4$ in $W^++8,9$jet final states.
    Green squares indicate results using the
    standard clustering procedure, while blue circles indicate results
    obtained from the winner-takes-all (WTA) approach as the number of jets
    exceeds six. See Sec.~\ref{sec:parton_shower} for details.}
  \label{fig:timing}
\end{figure}

Technically, the merging algorithm described above involves multiple stages:
\begin{enumerate}
\item\label{step:me} The computation of fixed-order results
\item The clustering and $\alpha_s$ reweighting
\item The parton shower and Sudakov reweighting
\end{enumerate}
In the past, different implementations have combined these steps in different
ways. Traditionally, the Sherpa event generator performs the jet clustering
during the computation of the fixed-order result and optimizes the Monte-Carlo
integrator based on the hard matrix element, including $\alpha_s$ reweighting.
The Pythia event generator relies on external matrix element providers~\cite{Alwall:2006yp}
to compute the perturbative inputs, and therefore a natural separation of Step~\ref{step:me}
from the remainder of the calculation occurs. We argue that this also provides the
more natural separation for improved compute performance. The reasons are twofold:
\begin{enumerate}
\item The parton shower and the clustering are probabilistic, in the sense that
  the number of particles produced in the shower, or the path chosen in the clustering
  are not known a priori. In contrast, the fixed-order perturbative calculations
  used as an input to the parton shower operate at fixed particle multiplicity,
  and always evaluate the same Feynman diagrams. By separating these two domains,
  we divide the program into two components with different program flow.
\item The computation of fixed-order results is very cumbersome at high multiplicity,
  even when making use of recursion relations. The corresponding unweighting
  efficiencies are usually very small. By comparison, both the parton shower and
  the jet clustering procedure are fast and consume significantly less memory.
  This is exemplified in Fig.~\ref{fig:timing}.
  Separating the two domains and storing results of the fixed-order calculation
  into intermediate event files allows to reuse the computationally most expensive
  parts of the simulation for calculations with different parton-shower or
  hadronization parameters.
\end{enumerate}
In the following we will discuss the problems related to fixed-order
calculations and parton-shower simulations on HPC architectures
in more detail, and present solutions that allow us to carry out
simulations relevant for the high-luminosity LHC.

\subsection{Fixed-order Computations}
\label{sec:fixed_order}
Fixed-order computations usually proceed in two steps: The optimization
stage uses adaptive algorithms like Vegas~\cite{Lepage:1977sw},
and possibly multi channels~\cite{Kleiss:1994qy}, to better
approximate the integrand, and therefore reduce the variance of the
Monte-Carlo integral that is evaluated. During the integration or event generation
stage, the integral is determined at high precision and
weighted or unweighted events are generated, and possibly stored.

Both stages are ideally suited for MPI parallelization.
During optimization, a large number of phase-space points must usually be
generated to provide the input for adaptive algorithms. MPI parallelization
can be achieved by independently producing a fraction of these points
on each MPI rank and subsequently collecting the results. This process
is repeated as needed. The computation of matrix elements can also be
thread-parallelized using for example the techniques
in~\cite{Gleisberg:2008fv,Giele:2010ks,Campbell:2015qma}.
During the event generation stage, any Monte-Carlo simulation is trivially
parallelizable. The only complication are I/O operations related to the
read-in of information related to the construction of the
hard matrix elements and to the parameters of the adaptive integrator.

The related files can be large at high particle multiplicity, because
the multi-channel integrator consists of many individual channels
(typically one per diagram). We choose to store this information in
a zip file by means of the libzippp interface~\cite{libzippp}.
We have implemented a parallelization layer into Comix, which both
maps the file access through libzippp to a std::i/ostream, and enables
read/write access to the actual zip file only on the master rank of the
MPI executable. During the read operation, the file content obtained by the
master rank is then broadcast to all ranks via the MPI Bcast function.
We have implemented the same technology into the LHAPDF6 parton distribution
library~\cite{Buckley:2014ana}.

\begin{figure}
  \includegraphics[width=\linewidth]{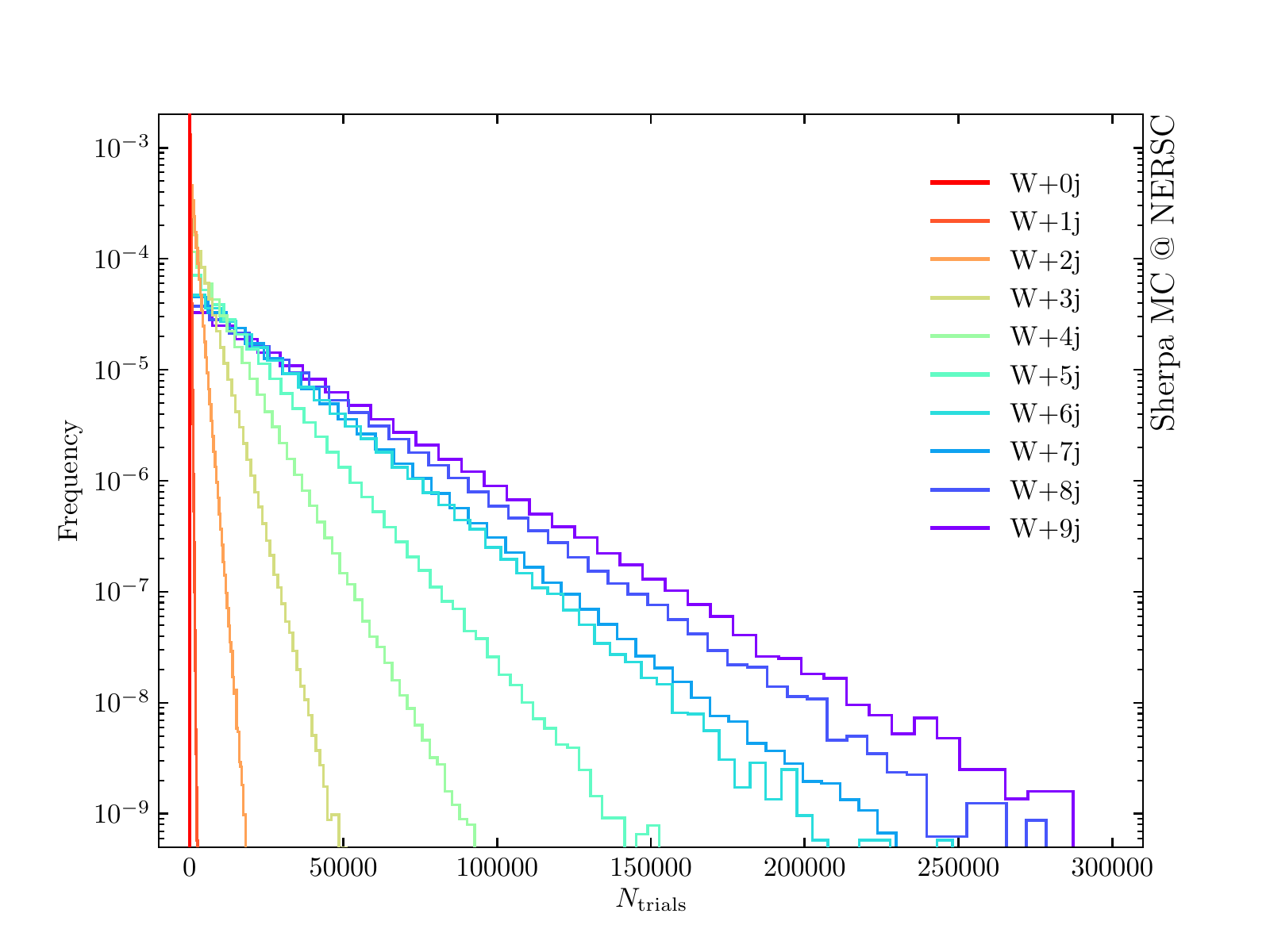}
  \caption{Distribution of the number of trials in the unweighting
    of $pp\to W^++nj$ at the 14~TeV LHC. Jets are required to have
    $p_{T,j}>20$~GeV and $|y_j|<6$.}
  \label{fig:number_of_trials}
\end{figure}
A significant complication occurs in the generation of unweighted events,
which is most easily explained by using an example. Consider
Fig.~\ref{fig:number_of_trials}, which shows the distribution of the
number of trials in the unweighting of $pp\to W^++nj$ events at
leading order QCD with Comix. Let us assume that we want to generate
$10^4$ $W+8j$ events. Generating them on a single rank, the timing
per event will be determined by the average number of trials, which
is given by the slope of the distribution. A trivial MPI parallelization
strategy would be to generate $10^4/N$ events on each rank for $N$
total ranks in the MPI run. This approach is bound to fail at large
$N$, as can be seen easily: Assume that $N=10^4$, and that one MPI
rank generates an event with 250000 trials. The timing of the overall
MPI run is then set by this rank, and until it has finished the remaining ranks are in an effective wait state. Therefore, the overall
event generation time is $10^4$ times the generation time for an
event with 250000 trials. Generally, for an executable parallelized
in this manner, the event generation time is set by the most
inefficient rank. In order to efficiently parallelize the simulation,
we might instead choose to generate weighted events and perform the
unweighting independently. While the timing will be uniform in this case,
weighted events require a large amount of disk storage, especially
if the unweighting efficiency is as low as in the eight jet case.
This holds true even if we store only the event weight. 

One may attempt to solve this problem by a hybrid approach:
Generate unweighted events, but limit the number of trials
to a fixed number $N_{\rm trial,max}$ that can be defined by the user
of the program and should be adapted to the efficiency and timing
for any given partonic final state. If the program reaches
$N_{\rm trial,max}$, and no event has passed the unweighting,
an event of weight zero is returned in order to achieve more
uniform timing. However, for a large number of nodes this approach
is still insufficient. We have verified this by disabling
the writeout of events\footnote{
  We would not be able to compare the scaling if we were to write out
  weighted events, due to the high storage requirements in the weighted
  event case. Disabling the writeout for unweighted events instead leaves
  the scaling behavior of the unweighted event generation intact.}
and comparing to weighted event generation, which then scales ideally.
Based on this finding, we implement a technique similar to Alpgen~\cite{Mangano:2002ea},
which was adapted to scale to over a million threads in~\cite{Childers:2015tyv}:
We generate a fixed number of weighted events, unweight on the fly,
and store only those events which pass the unweighting procedure.
This leaves us with a variable number of unweighted events, which cannot
be predicted, as it depends on the efficiency of the integrator and
is subject to statistical fluctuations. However, the scaling behavior
is determined solely by the I/O performance (cf.\ Sec.~\ref{sec:results}).

To reduce the memory footprint of the executable, we allow to manually
partition the partonic processes generated by Comix into sets of
irreducible groups with identical diagrammatic structure of the
hard matrix elements.

\subsection{Parton Showers and Merging}
\label{sec:parton_shower}
The simulation of parton showers and the associated jet clustering
procedure behave intrinsically different from the fixed-order
computations discussed above, and they involve their own, unique
computational challenges. Since the parton shower is a Markov Chain
Monte Carlo, the particle multiplicity and the flavor structure varies
event by event, such that memory usage and program flow cannot
be predicted. Since the typical parton multiplicity is
$\mathcal{O}(10-100)$, and only one possible parton-shower history
is generated per event, this does not present a problem.
The jet clustering procedure as the inverse operation to the parton shower
does however create a challenge:
In order to determine the correct probability for any given history,
one must construct \emph{all} histories corresponding to a given parton-level
final state in the fixed-order calculation~\cite{Andre:1997vh}.
For indistinguishable final-state particles, the number of possible
histories grows at least factorially, depending on the parton-shower
recoil scheme. Both timing and memory usage of the executable therefore
increase rapidly with the final-state multiplicity.
This is exemplified in Fig.~\ref{fig:timing} for the exclusive
contributions to a $W^++$jets calculation in a multi-jet merged approach.
The computational complexity increases by approximately a factor~1.5
for each additional final-state jet.\footnote{Note that
  the scaling for $N>6$ is improved by the winner takes it all
  approach explained below.}
As argued in Sec.~\ref{sec:fixed_order} any non-uniform timing
in the calculation will eventually break the scaling, even if the
simulation is trivially parallelizable, as in the case of parton showers.
Starting with $pp\to W/Z+7$jet final states, the jet clustering procedure
also begins to exhaust the memory of modern computers (at $\approx$4~GB/core).

These problems must be tackled by changes to the underlying merging algorithms.
We use the CKKW-L technique~\cite{Lonnblad:2001iq}, and amend it as follows:
In order to make the generation of parton-level events independent
of the merging procedure, we use a standard $k_T$ jet algorithm
to regularize the hard matrix elements and to veto the parton shower.
In parton-level configurations that exceed the complexity of a
$W/Z+6$jet final state, we perform the first clustering steps using the
``winner takes it all'' (WTA) approach: The clustering with largest
probability is chosen, independent of the parton-shower history of
the clustered lower-multiplicity state.
This procedure is repeated until the final state has been reduced by the
jet clustering to a $W/Z+6$jet configuration. At this point, the standard
algorithm resumes. We have verified that the change to physical cross
sections and distributions arising from this WTA approach is at the level
of a few percent, and is significantly lower than the renormalization and
factorization scale uncertainties arising from the perturbative expansion.
Event samples with different final-state multiplicities at parton level
are processed independently, in order to make the timing of the clustering
and subsequent parton showering as uniform as possible.
We use ASCR's DIY framework~\cite{peterka_ldav11,morozov_ldav16}
to parallelize the particle-level event simulation.

\section{New Framework for Massively Parallel Processing}
\label{sec:framework}
It remains to discuss the combination of the fixed-order and parton shower
calculations described above into a consistent, multi-jet merged event simulation.
This relies on the efficient communication of event properties at the parton level
by means of intermediate event files that are similar in spirit to Les Houches
Event Files (LHEF)~\cite{Alwall:2006yp}. The LHEF standard is widely used in the
high-energy physics community. It is based on XML, which poses a challenge 
for I/O operations, in particular the simultaneous read access when processing
event information in heavily parallelized workflows. We address this problem
by proposing a new format based on HDF5~\cite{HDF5}, which is designed specifically
for processing large amounts of data on HPC machines. HDF5 uses a computing model
not too dissimilar from databases. The data stored is organized in data sets,
that can be thought of as tables of standard types such as integer and float.
These data sets can be organized in groups in order to create hierarchical structures.
We strive to keep the new HDF5 event file standard as similar to the LHEF standard
as possible.

The LHEF format comprises global properties as well as event-wise
properties~\cite{Alwall:2006yp}. Global properties include process information
(i.e.\ the type of collisions), total cross-sections as well as reweighting information.
The event-wise properties are the process ID, the event weight, the scale of the
hard process as well as the values of $\alpha_\text{QCD}$, $\alpha_\text{QED}$
and of course the list of particles generated.  The latter contain the momentum four-vectors,
information on particle identification (charges, spin, lifetime) as well as their genealogy.
We propose to organize the HDF5 structure of Les Houches events such that each 
quantity normally stored in an event-wise block of XML is instead written
to independent HDF datasets. We suggest to have separate groups for global, 
event-wise and particle properties.

\begin{table}[t]
    \centering
    \begin{tabular}{lc}
        Dataset & data type \\ \toprule
    {\sc PDFgroupA} & int \\
    {\sc PDFgroupB} & int \\
    {\sc PDFsetA} & int \\
    {\sc PDFsetB} & int \\
    {\sc beamA}  & int \\
    {\sc beamB} & int \\
    {\sc energyA} & double \\
    {\sc energyB} & double \\
    {\sc numProcesses} & int \\
    {\sc weightingStrategy} & int \\
    \end{tabular}
    \caption{Datasets and data types in the {\it init} group.}
    \label{tab:group-init}
\end{table}
\begin{table}[t]
    \centering
    \begin{tabular}{lc}
        Dataset & data type \\ \toprule
    {\sc procId}     & int \\\bottomrule
    {\sc xSection}   & double\\
    {\sc error}      & double\\
    {\sc unitWeight} & double\\
    {\sc npLO} & int\\
    {\sc npNLO} & int \\
    \end{tabular}
    \caption{Datasets and data types in the {\it procInfo} group.}
    \label{tab:proc-info}
\end{table}

\subsubsection{Global properties}
We follow the general idea of the XML-tags used in LHE files and define
a group {\it{init}}. The names of the data sets as well as their data types
are summarized in Tab.~\ref{tab:group-init}. Properties of the individual
processes are stored in the group {\it procInfo}. The names of the data sets
and their data types are summarized in Tab.~\ref{tab:proc-info}.

\subsubsection{Event-wise properties}
The information that is unique to a single event (except its particles)
is stored in the {\it event} group. Table~\ref{tab:group-event} gives
an overview of the data set name and data types used.

\begin{table}[t]
    \centering
    \begin{tabular}{lc}
        Dataset & data type \\ \toprule
        {\sc nparticles}  & int    \\
        {\sc start}          & int \\
        {\sc pid}         & int    \\
        {\sc weight}      & double \\
        {\sc scale}       & double \\
        {\sc fscale}       & double \\
        {\sc rscale}      & double \\
        {\sc aqed}        & double \\
        {\sc aqcd}       & double \\ \bottomrule
        {\bf{\sc npLO}}       & int \\
        {\bf{\sc npNLO}}      & int \\
        {\sc trials}          & double
    \end{tabular}
    \caption{\label{tab:group-event} Data sets and data types in the {\it event} group.}
\end{table}

The number {\it start} identifies the location
of the first record in the additional group named {\it particle},
which stores information about individual particles that belong to
each event. A summary of the information stored in the {\it particle}
group is given in Tab.~\ref{tab:group-particle}.

\begin{table}[t]
    \centering
    \begin{tabular}{lc}
        Dataset & data type \\ \toprule
        {\sc id}       & int    \\
        {\sc status}   & int    \\
        {\sc mother1}  & int    \\
        {\sc mother2}  & int    \\
        {\sc color1}   & int    \\
        {\sc color2}   & int    \\
        {\sc px}       & double \\
        {\sc py}       & double \\
        {\sc pz}       & double \\
        {\sc e}        & double \\
        {\sc m}        & double \\
        {\sc lifetime} & double \\
        {\sc spin}     & double \\ \bottomrule
    \end{tabular}
    \caption{\label{tab:group-particle} Data sets and data types in the {\it particle} group.}
\end{table}

\section{Numerical Results}
\label{sec:results}
In this section we present first phenomenological results
generated with the new event generation framework and discuss
its computing performance. We consider proton-proton collisions
at the high-luminosity LHC at $\sqrt{s}=14~{\rm TeV}$. We use
the CT14 NNLO PDF set~\cite{Dulat:2015mca} and define the strong
coupling accordingly. Our modified parton-level event generator
is based on Comix~\cite{Gleisberg:2008fv} as included in Sherpa
version 2.2.4~\cite{Gleisberg:2008ta}. Our modified particle-level
event generator, is based on Pythia~8~\cite{Sjostrand:2014zea} and will be part of the next Pythia release.

\begin{table*}[t]
    \centering
    \begin{tabular}{l|C{13mm}C{13mm}C{13mm}C{13mm}C{13mm}C{13mm}C{13mm}C{13mm}C{13mm}C{13mm}}
         $pp\to X+n\;{\rm jets}$ & \multicolumn{10}{c}{cross section [pb]} \\[1mm]
         $X$ / $n$ & 0 & 1 & 2 & 3 & 4 & 5 & 6 & 7 & 8 & 9 \\\hline
         $W^+$\vphantom{$\int_A^{B^C}$} & 9908(29) & 2523(8) & 1067(7) & 404(4) & 148(1) & 49.3(5) & 15.8(2) & 5.2(2) & 1.30(8) & 0.330(6) \\
         $W^-$\vphantom{$\int_A^B$} & 7496(21) & 1898(6) & 760(4) & 278(2) & 94(1) & 29.8(3) & 9.29(9) & 2.71(7) & 0.63(2) & 0.170(3) \\
         $Z$ \vphantom{$\int_A^B$} & 1661(3) & 464(1) & 193.6(8) & 72.2(3) & 25.7(2) & 8.61(8) & 2.74(3) & 0.82(2) & 0.211(3) & 0.057(1)
    \end{tabular}
    \caption{Inclusive cross sections at the LHC at $\sqrt{s}=14~{\rm TeV}$
    using the CT14nnlo PDF set and a correspondingly defined strong coupling.
    Jets are defined using the $k_T$ clustering algorithm with $R=0.4$, $p_{T,j}>20\;{\rm GeV}$ and $|\eta_j|<6$.}
    \label{tab:partonic_xs}
\end{table*}
Table~\ref{tab:partonic_xs} shows the parton-level cross sections
for the event samples produced by Comix and used in the merging.
Jets are defined using the $k_T$ clustering algorithm with $R=0.4$,
$p_{T,j}>20\;{\rm GeV}$ and $|\eta_j|<6$. Following the good agreement
between parton-level and particle-level results established
in~\cite{Bellm:2019yyh}, and the good agreement between fixed-order
and MINLO~\cite{Hamilton:2012np} results established in~\cite{Anger:2017nkq},
the renormalization and factorization scales are set to $\hat{H}_T'/2$~\cite{Berger:2010zx}.


\begin{figure}[t]
    \centering
    \includegraphics[width=\linewidth]{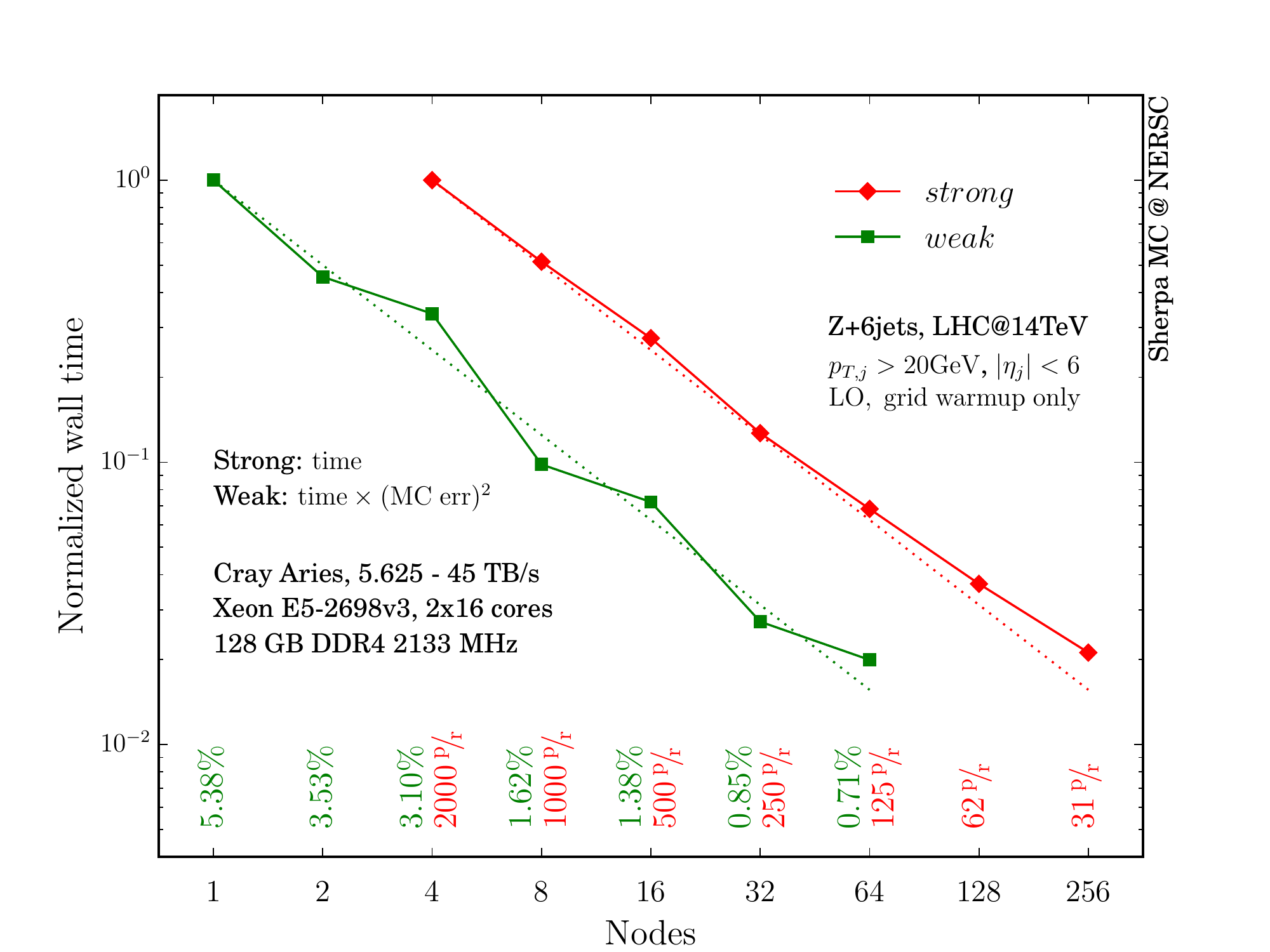}
    \caption{Test of scaling behavior in the optimization stage
      of the parton-level calculation. In the strong scaling test
      (red) the number of points per rank ($^p\!/\!_r$) is
      inversely proportional to the number of ranks. In the weak
      scaling test (green), $^p\!/\!_r$ is constant and the computation
      time is multiplied by the square of the Monte-Carlo error.}
    \label{fig:int_scaling}
\end{figure}
Figure~\ref{fig:int_scaling} shows the scaling behavior of the
parton-level calculation during the optimization stage.
We limit the number of quarks to $\le6$ in $W^++6,7$jet
and to $\le4$ in $W^++8,9$jet final states. The red
line corresponds to a test of strong scaling. The number of
non-zero points generated per rank and optimization step, which we
denote as $^p\!/\!_r$, is inversely proportional to the number of ranks.
The results deviate from the perfect scaling assumption shown
by the red dotted line for fewer than $100\,^p\!/\!_r$. This is
somewhat expected, as cut efficiencies tend to be non-uniform
across various ranks when the number of points per rank is too low.
The problem could in principle be addressed by not requesting
a fixed number of non-zero phase-space points per optimization step,
but by requesting a fixed number of points (zero and non-zero).
However, such a procedure would not guarantee that the optimization
can be carried out efficiently, because the adaptive Monte-Carlo
integration depends on sufficient statistics~\cite{Lepage:1977sw,Kleiss:1994qy}.
Fluctuations in the number of terms to be evaluated in the
Monte-Carlo integration over color~\cite{Maltoni:2002mq,Duhr:2006iq}
also contribute to the non-uniform timing, and therefore to the
breakdown of scaling when the number of points per rank is too low.
In the test of weak scaling, we keep the number of points per rank
a constant and instead multiply the computation time by the square
of the Monte-Carlo error. For a fixed number of non-zero points
per rank, the error scales as $1/\sqrt{N}$, with $N$ being the number of ranks.
This behavior is what we observe on average, while the randomness
in the Monte-Carlo integration generates fluctuations around the
projected timing for ideal scaling. We remark, however, that the
weak scaling test is rather academic. For a calculation like
$W/Z+9$jets, the timing of the overall Monte-Carlo integration
is limited by the allocated computing time. The job shape
as well as the number of points per rank need to be determined
such that the wall time of the calculation can be minimized.

\begin{figure}[t]
    \centering
    \includegraphics[width=\linewidth]{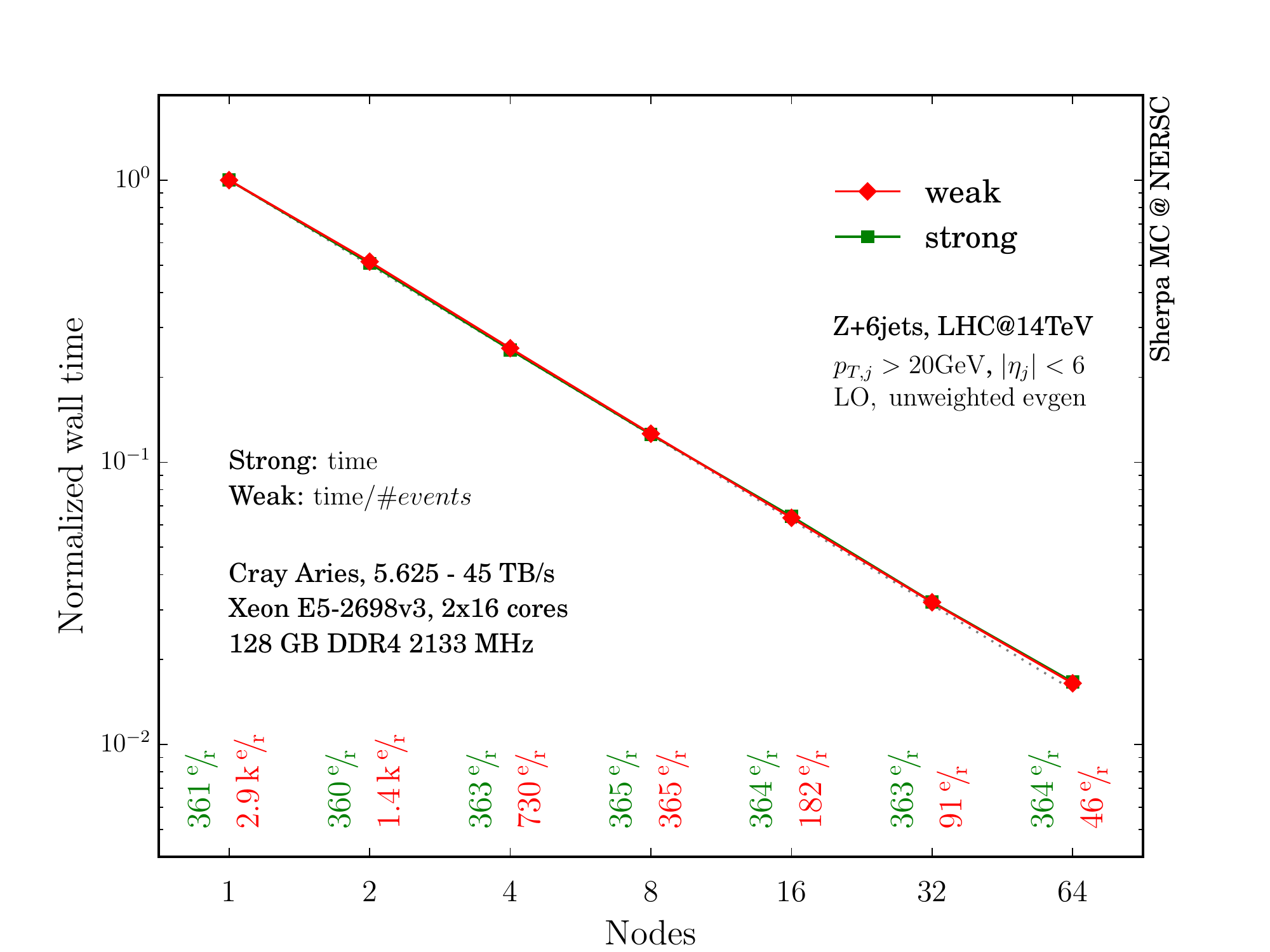}
    \caption{Test of scaling behavior in the event generation stage
      of the parton-level calculation. In the strong scaling test (red)
      the number of events per rank, $^e\!/\!_r$, is inversely proportional
      to the number of ranks. In the weak scaling test (green)
      the computation time is divided by the number of events.}
    \label{fig:gen_scaling}
\end{figure}
Figure~\ref{fig:gen_scaling} displays the scaling behavior of the parton-level
event generation. Both the strong (red) and the weak (green) scaling test show
satisfactory performance up to 2048 cores (64 nodes) on the Cori system at
NERSC~\cite{cori}\footnote{In order to speed up the writeout of events,
  we have made use of the Burst Buffer.}
While the scaling behavior is much better than when requesting a fixed number
of unweighted events (cf.\ the discussion in Sec.~\ref{sec:framework}),
it is not yet ideal, because the file size for event storage is determined
dynamically. Optimizing the HDF5 output parameters may lead to further
improvements.

\begin{figure}[t]
    \centering
    \includegraphics[width=\linewidth]{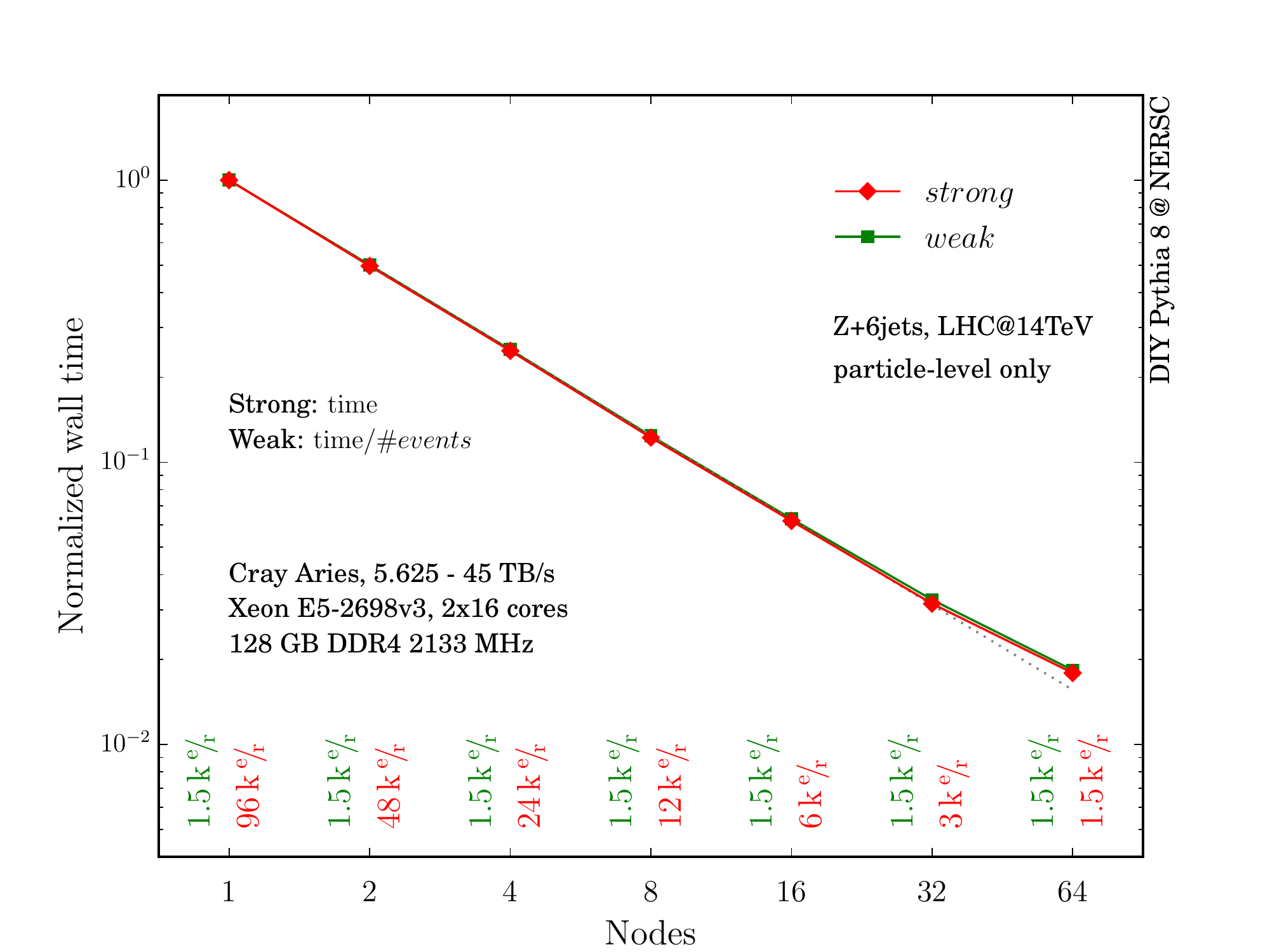}
    \caption{Test of scaling behavior in the particle-level simulation.
      In the strong scaling test (red) the number of events per rank, $^e\!/\!_r$,
      is inversely proportional to the number of ranks. In the weak scaling test (green)
      the computation time is divided by the number of events.}
    \label{fig:pl_scaling}
\end{figure}
Figure~\ref{fig:pl_scaling} shows the scaling behavior in the particle-level
event simulation with Pythia~8~\cite{Sjostrand:2014zea}
and DIY~\cite{morozov_ldav16}.
Both the strong (red) and the weak (green) scaling test show a deviation
from the ideal pattern, which can be attributed to the I/O overhead.
In this context it is important to note that we have set the number of events
for the weak and strong scaling test to be the same at 2048 MPI ranks (64 nodes).
This implies a larger I/O overhead in the weak scaling test and explains
the slightly worse scaling behavior at smaller number of ranks.
It exemplifies that even for relatively complex simulations, like $Z+6$jets,
the I/O overhead quickly begins to dominate.

\begin{figure}[t]
    \centering
    \includegraphics[width=\linewidth]{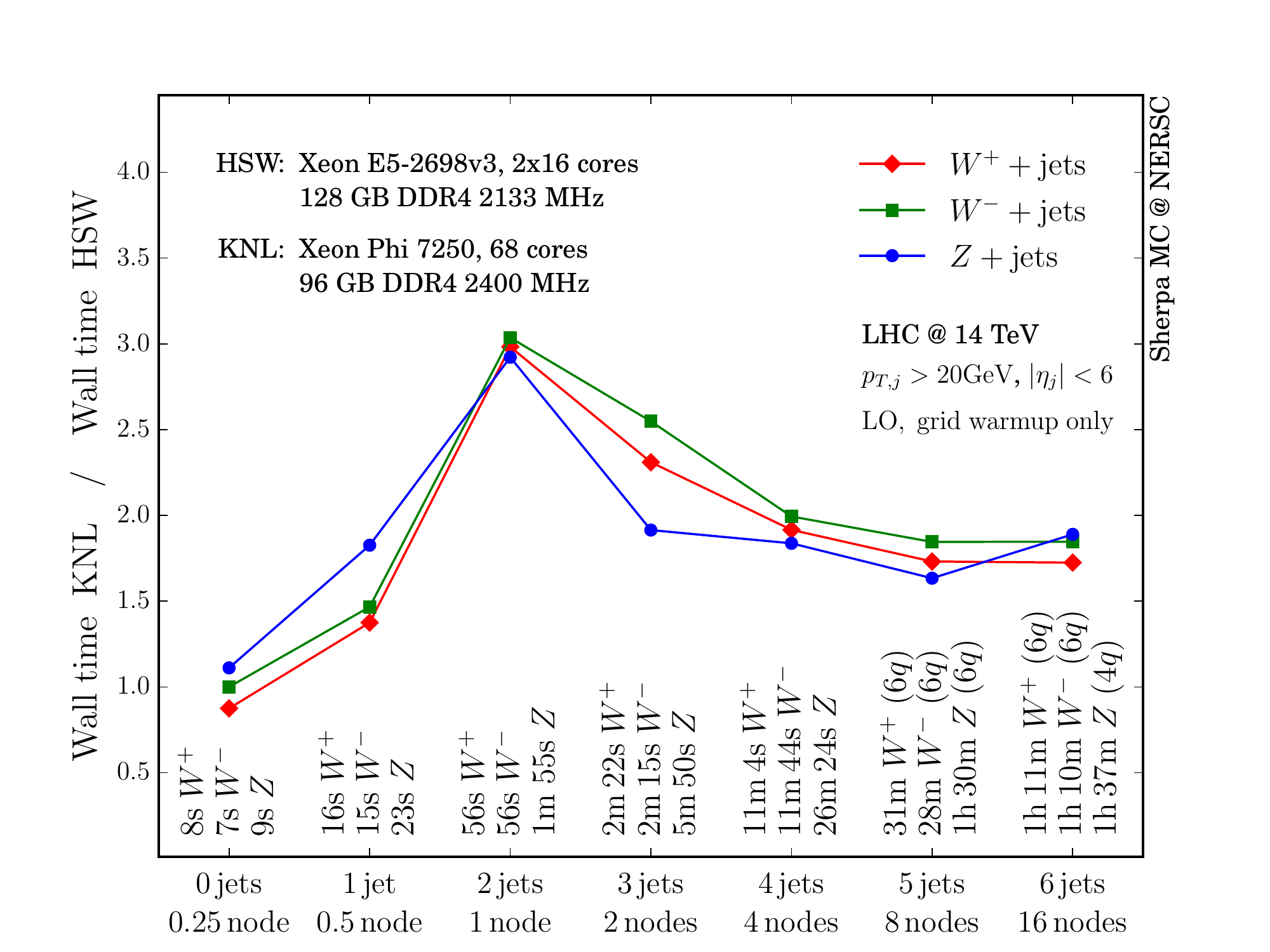}
    \caption{Performance comparison of the parton-level event generation
      framework during the optimization stage between the KNL and Haswell
      architectures of Cori at NERSC~\cite{cori}.}
    \label{fig:knl_vs_hsw}
\end{figure}
Finally, in Fig.~\ref{fig:knl_vs_hsw} we show a test of our new parton-level
event generation framework during the optimization stage on the KNL nodes
of Cori Phase~II at NERSC~\cite{cori}.
The memory footprint of the executable needs to be kept at a minimum for this test,
in order not to exceed the 1.4~GB RAM/core available on this architecture. We have
exhausted all four hyperthreads per core through MPI, which reduces the available
memory per rank to 350~MB. The performance is reduced compared to a Haswell
architecture by a factor of two to three, depending on the jet multiplicity.
Given the overall computing time needed for the optimization, this presents
more an academic than a practical problem at this stage.

\begin{figure}[t]
    \centering
    \includegraphics[width=\linewidth]{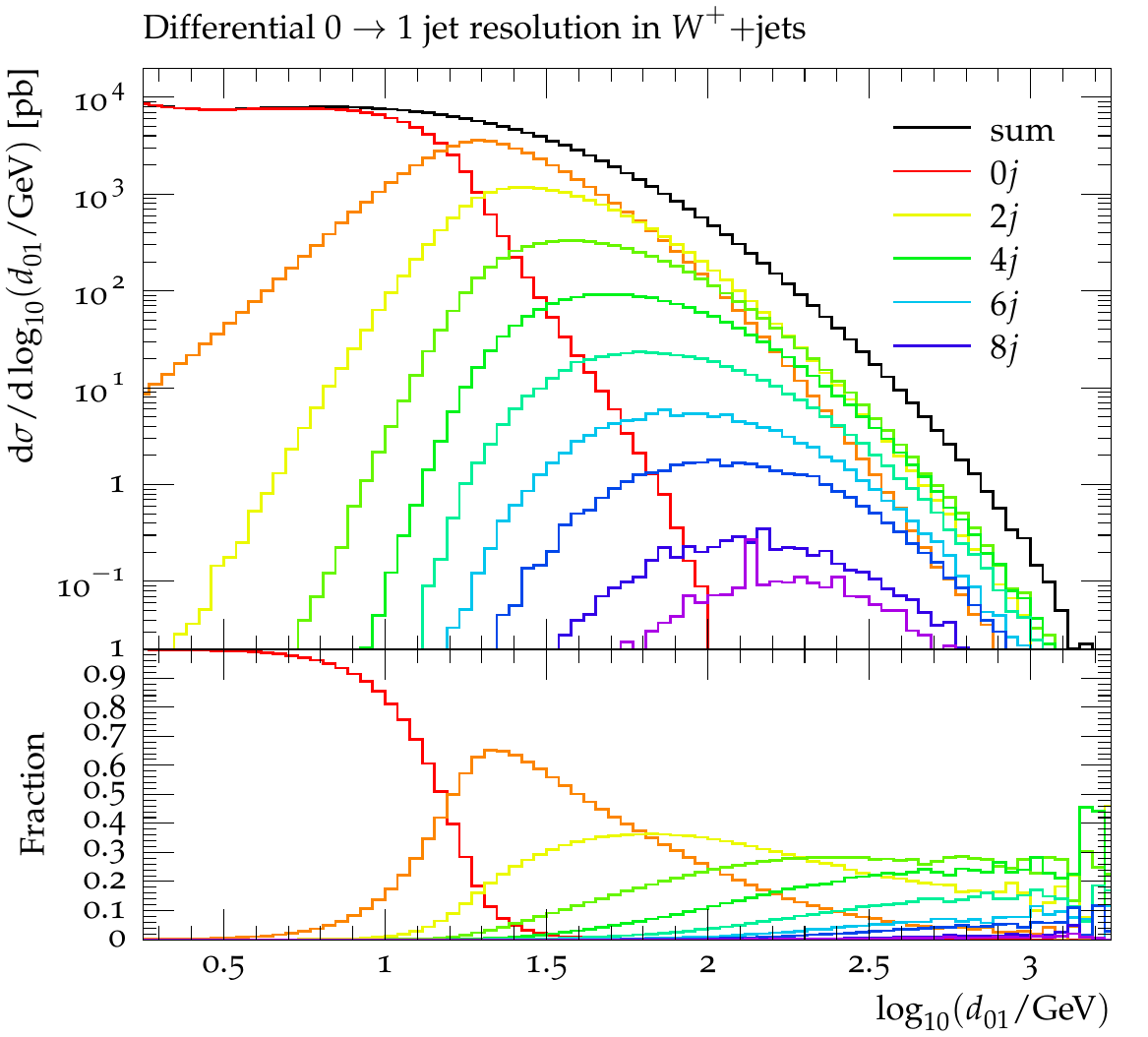}
    \includegraphics[width=\linewidth]{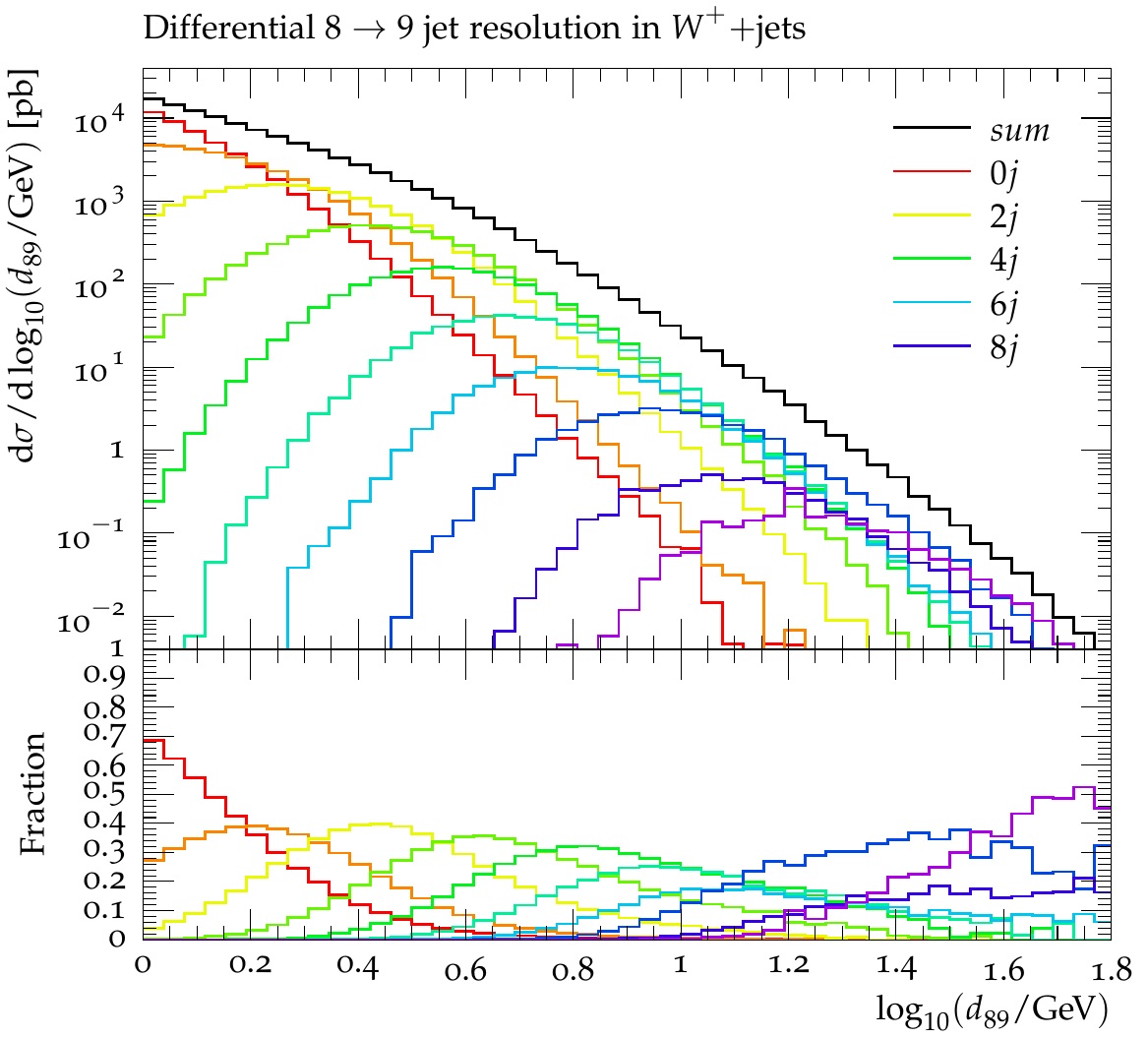}
    \caption{Differential $0\to 1$ and $8\to 9$ $k_T$ jet rates in $Z$+jets events.
      Colored lines represent the $Z+0$-$9$ jet contributions in the merging.}
    \label{fig:jetrates}
\end{figure}
Figure~\ref{fig:jetrates} shows differential jet rates in $Z+$jets events
using the $k_T$ algorithm~\cite{Catani:1993hr}. The colored lines represent the
contributions from event samples of different jet multiplicity at parton level.
It is interesting to note that high multiplicity configurations play
a significant role in the $0\to 1$ jet rate at high $p_T$.

\begin{figure*}[t]
    \centering
    \begin{minipage}{0.475\textwidth}
      \includegraphics[width=\linewidth]{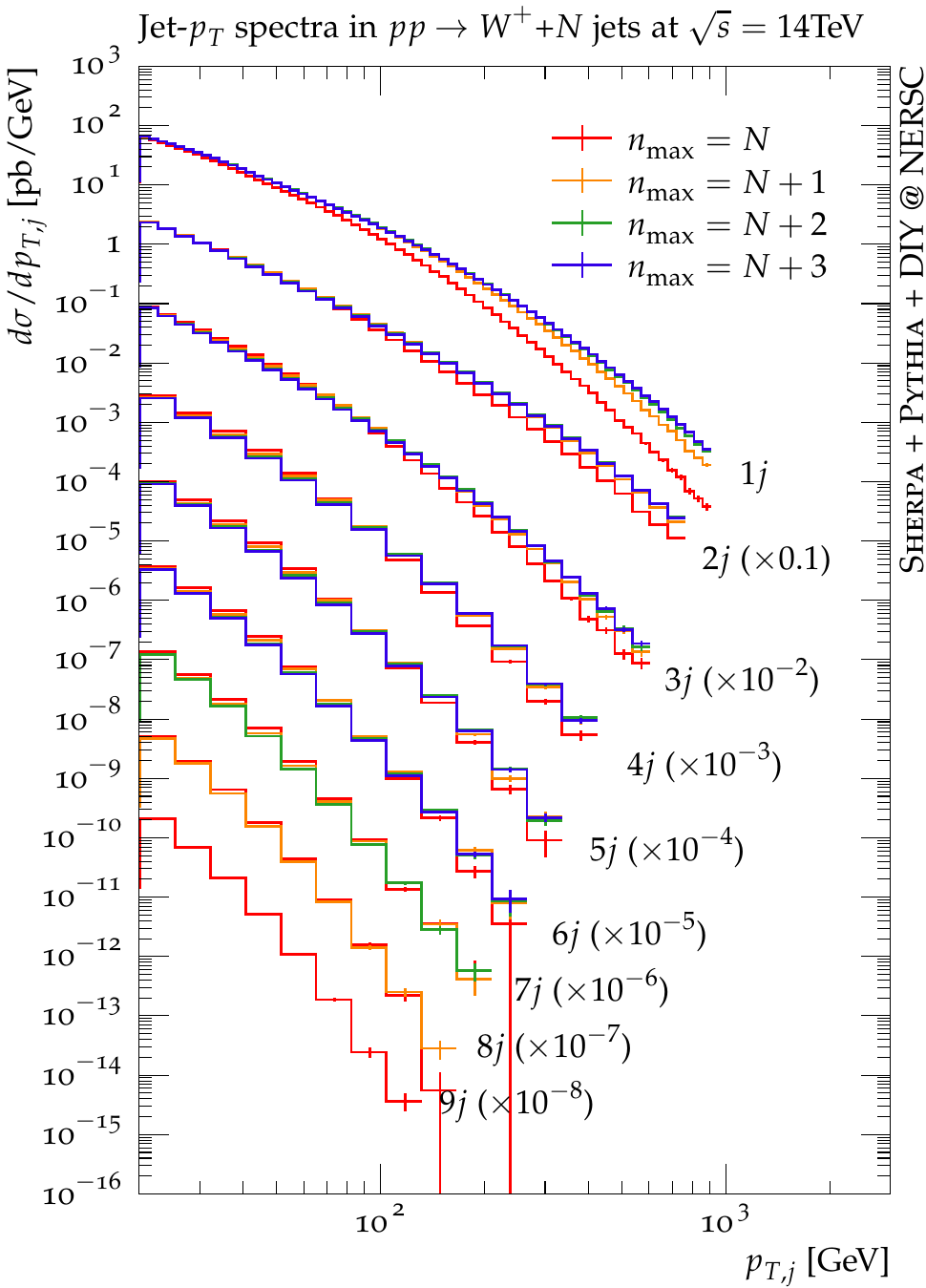}
    \end{minipage}\hfill
    \begin{minipage}{0.475\textwidth}
      \includegraphics[width=\textwidth]{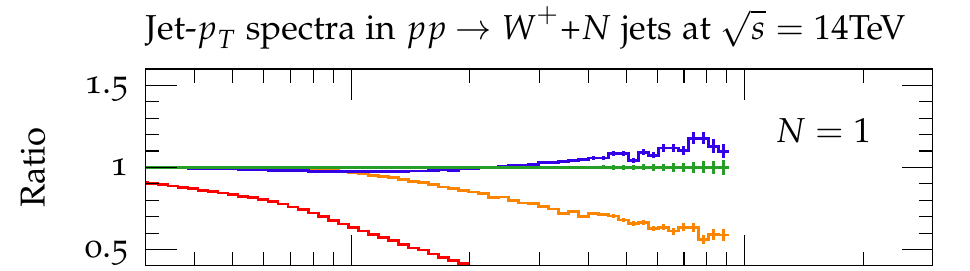}\\[-1mm]
      \includegraphics[width=\textwidth]{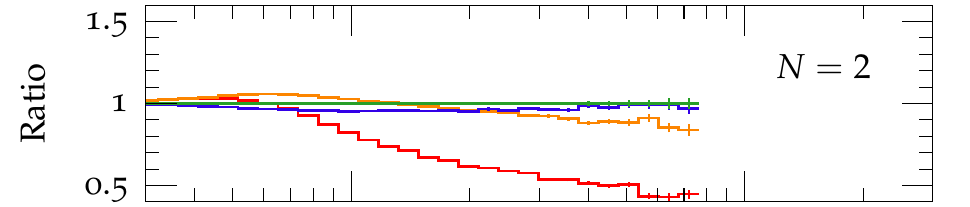}\\[-1mm]
      \includegraphics[width=\textwidth]{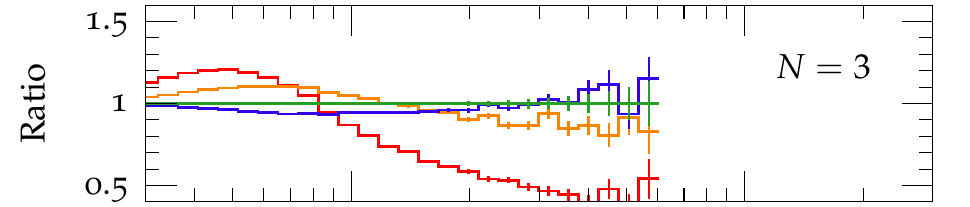}\\[-1mm]
      \includegraphics[width=\textwidth]{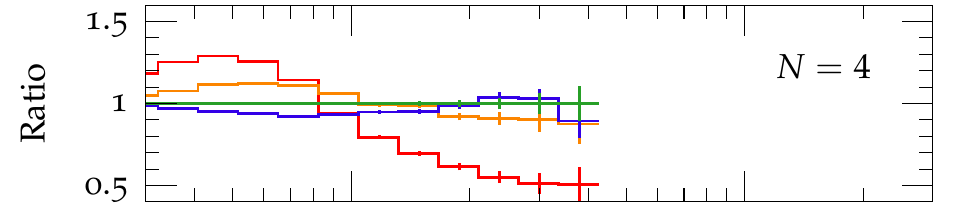}\\[-1mm]
      \includegraphics[width=\textwidth]{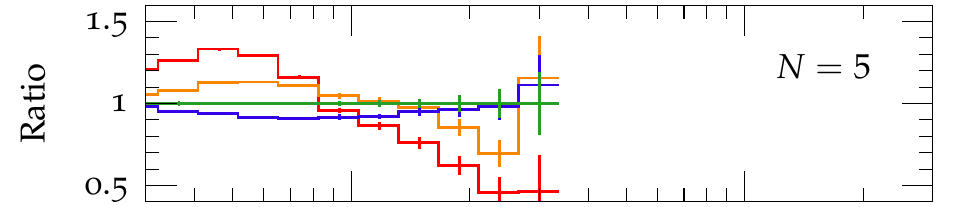}\\[-1mm]
      \includegraphics[width=\textwidth]{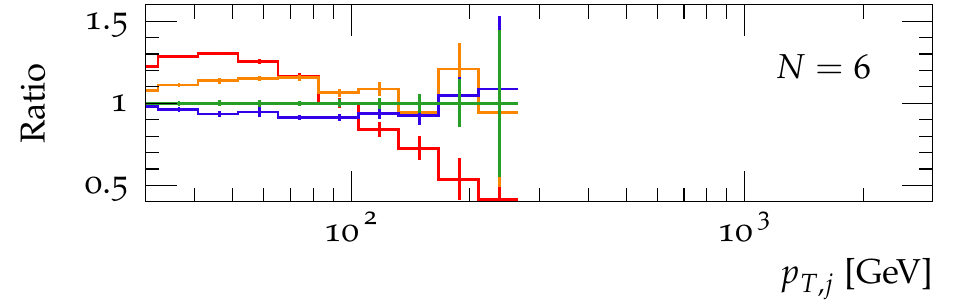}
    \end{minipage}
    \caption{Jet transverse momentum distributions in $W^+$+jets events.
      We show a comparison of multi-jet merged simulations where the
      maximum jet multiplicity, $n_{\rm max}$, is set to the number of
      measured jets, $N$ (red), to $N+1$ (green), $N+2$ (blue) and
      $N+3$ (purple).}
    \label{fig:jet_pT}
\end{figure*}
Figure~\ref{fig:jet_pT} shows the jet transverse momenta for a varying
maximum jet multiplicity, $n_{\rm max}$, starting at the number of
observed jets, $N$, and ranging up to $N+3$ (if possible). As observed
in~\cite{Krauss:2004bs,Krauss:2005nu}, the predictions for a low number
of jets, $N$, stabilize at $n_{\rm max}=N+1$. It is interesting
to observe that the correction to the $n_{\rm max}=N$ sample for $N\ge 3$
is negative at low $p_T$, indicating that the default parton-shower
tune of Pythia 8.240 overestimates the emission rate in this configuration.

\section{Conclusions}
\label{sec:conclusions}
We have presented a new event generation framework, suitable for massively
parallel processing of multi-jet merged simulations for current and future
collider experiments. Making use of high performance computing resources,
we have computed the first predictions for single vector boson production
in proton proton collisions with up to nine jets described at leading order
accuracy. This presents a significant improvement of the currently available
simulations, which are capable of reaching leading order accuracy for only
up to six jets in the final state. The results of our parton-level event
generation campaign can be obtained from~\cite{schulz_holger_2019_2678039,
  schulz_holger_2019_2678055,schulz_holger_2019_2678091}
and can be used as an input to particle-level simulations in experimental
analyses as well as phenomenology.
The modified parton-level and particle-level event generators,
together with wrapper scripts and setups needed to perform
similar simulations, are publicly available~\cite{hpcgenrepo}.

We have shown that the algorithms employed in our simulation scale as far
as can reasonably be expected, given the problem size of the computation.
Limitations arise during the integration stage from the limited number
of phase-space points that are generated in each integration step, and
in the event generation stage from the I/O limitations of the host system.
We achieve best performance with MPI parallelization on up to 2048 ranks
(64 nodes) of the Cori system at NERSC~\cite{cori}.
\vspace*{2mm}

\section*{Acknowledgments}
We thank our colleagues in the Sherpa and Pythia collaborations, in
particular Frank Siegert and Stephen Mrenna, for discussions and support.
We are grateful to Marc Paterno, Jim Kowalkowski, Chris Green and
Saba Sehrisch for help with HDF5. We thank Rick and the team behind
HDFql for their support.
We thank Taylor Childers for numerous fruitful discussions
on code performance and for comments on the manuscript.
This material is based upon work supported by the U.S. Department of Energy,
Office of Science, Office of Advanced Scientific Computing Research,
Scientific Discovery through Advanced Computing (SciDAC) program,
grants ``HEP Data Analytics on HPC'', No.~1013935 and
``HPC framework for event generation at colliders''. 
It was supported by the U.S. Department of Energy under contracts
DE-AC02-76SF00515 and DE-AC02-07CH11359 and used resources of the
National Energy Research Scientific Computing Center (NERSC),
a U.S. Department of Energy Office of Science User Facility
operated under Contract No. DE-AC02-05CH11231.

\bibliography{journal}

\begin{thebibliography}{80}
\expandafter\ifx\csname natexlab\endcsname\relax\def\natexlab#1{#1}\fi
\expandafter\ifx\csname bibnamefont\endcsname\relax
  \def\bibnamefont#1{#1}\fi
\expandafter\ifx\csname bibfnamefont\endcsname\relax
  \def\bibfnamefont#1{#1}\fi
\expandafter\ifx\csname citenamefont\endcsname\relax
  \def\citenamefont#1{#1}\fi
\expandafter\ifx\csname url\endcsname\relax
  \def\url#1{\texttt{#1}}\fi
\expandafter\ifx\csname urlprefix\endcsname\relax\def\urlprefix{URL }\fi
\providecommand{\bibinfo}[2]{#2}
\providecommand{\eprint}[2][]{\url{#2}}

\bibitem[{\citenamefont{Drell and Yan}(1970)}]{Drell:1970wh}
\bibinfo{author}{\bibfnamefont{S.~D.} \bibnamefont{Drell}} \bibnamefont{and}
  \bibinfo{author}{\bibfnamefont{T.-M.} \bibnamefont{Yan}},
  \bibinfo{journal}{Phys. Rev. Lett.} \textbf{\bibinfo{volume}{25}},
  \bibinfo{pages}{316} (\bibinfo{year}{1970}), \bibinfo{note}{[Erratum: Phys.
  Rev. Lett.25,902(1970)]}.

\bibitem[{\citenamefont{Drell and Yan}(1971)}]{Drell:1970yt}
\bibinfo{author}{\bibfnamefont{S.~D.} \bibnamefont{Drell}} \bibnamefont{and}
  \bibinfo{author}{\bibfnamefont{T.-M.} \bibnamefont{Yan}},
  \bibinfo{journal}{Annals Phys.} \textbf{\bibinfo{volume}{66}},
  \bibinfo{pages}{578} (\bibinfo{year}{1971}).

\bibitem[{\citenamefont{Altarelli et~al.}(1978)\citenamefont{Altarelli, Ellis,
  and Martinelli}}]{Altarelli:1978id}
\bibinfo{author}{\bibfnamefont{G.}~\bibnamefont{Altarelli}},
  \bibinfo{author}{\bibfnamefont{R.~K.} \bibnamefont{Ellis}}, \bibnamefont{and}
  \bibinfo{author}{\bibfnamefont{G.}~\bibnamefont{Martinelli}},
  \bibinfo{journal}{Nucl. Phys.} \textbf{\bibinfo{volume}{B143}},
  \bibinfo{pages}{521} (\bibinfo{year}{1978}), \bibinfo{note}{[Erratum: Nucl.
  Phys.B146,544(1978)]}.

\bibitem[{\citenamefont{Altarelli et~al.}(1979)\citenamefont{Altarelli, Ellis,
  and Martinelli}}]{Altarelli:1979ub}
\bibinfo{author}{\bibfnamefont{G.}~\bibnamefont{Altarelli}},
  \bibinfo{author}{\bibfnamefont{R.~K.} \bibnamefont{Ellis}}, \bibnamefont{and}
  \bibinfo{author}{\bibfnamefont{G.}~\bibnamefont{Martinelli}},
  \bibinfo{journal}{Nucl. Phys.} \textbf{\bibinfo{volume}{B157}},
  \bibinfo{pages}{461} (\bibinfo{year}{1979}).

\bibitem[{\citenamefont{Hamberg et~al.}(1991)\citenamefont{Hamberg, van
  Neerven, and Matsuura}}]{Hamberg:1990np}
\bibinfo{author}{\bibfnamefont{R.}~\bibnamefont{Hamberg}},
  \bibinfo{author}{\bibfnamefont{W.}~\bibnamefont{van Neerven}},
  \bibnamefont{and} \bibinfo{author}{\bibfnamefont{T.}~\bibnamefont{Matsuura}},
  \bibinfo{journal}{Nucl.Phys.} \textbf{\bibinfo{volume}{B359}},
  \bibinfo{pages}{343} (\bibinfo{year}{1991}).

\bibitem[{\citenamefont{van Neerven and Zijlstra}(1992)}]{vanNeerven:1991gh}
\bibinfo{author}{\bibfnamefont{W.~L.} \bibnamefont{van Neerven}}
  \bibnamefont{and} \bibinfo{author}{\bibfnamefont{E.~B.}
  \bibnamefont{Zijlstra}}, \bibinfo{journal}{Nucl. Phys.}
  \textbf{\bibinfo{volume}{B382}}, \bibinfo{pages}{11} (\bibinfo{year}{1992}),
  \bibinfo{note}{[Erratum: Nucl. Phys.B680,513(2004)]}.

\bibitem[{\citenamefont{Gehrmann-De~Ridder
  et~al.}(2016{\natexlab{a}})\citenamefont{Gehrmann-De~Ridder, Gehrmann,
  Glover, Huss, and Morgan}}]{Ridder:2015dxa}
\bibinfo{author}{\bibfnamefont{A.}~\bibnamefont{Gehrmann-De~Ridder}},
  \bibinfo{author}{\bibfnamefont{T.}~\bibnamefont{Gehrmann}},
  \bibinfo{author}{\bibfnamefont{E.~W.~N.} \bibnamefont{Glover}},
  \bibinfo{author}{\bibfnamefont{A.}~\bibnamefont{Huss}}, \bibnamefont{and}
  \bibinfo{author}{\bibfnamefont{T.~A.} \bibnamefont{Morgan}},
  \bibinfo{journal}{Phys. Rev. Lett.} \textbf{\bibinfo{volume}{117}},
  \bibinfo{pages}{022001} (\bibinfo{year}{2016}{\natexlab{a}}),
  \eprint{1507.02850}.

\bibitem[{\citenamefont{Boughezal
  et~al.}(2016{\natexlab{a}})\citenamefont{Boughezal, Campbell, Ellis, Focke,
  Giele, Liu, and Petriello}}]{Boughezal:2015ded}
\bibinfo{author}{\bibfnamefont{R.}~\bibnamefont{Boughezal}},
  \bibinfo{author}{\bibfnamefont{J.~M.} \bibnamefont{Campbell}},
  \bibinfo{author}{\bibfnamefont{R.~K.} \bibnamefont{Ellis}},
  \bibinfo{author}{\bibfnamefont{C.}~\bibnamefont{Focke}},
  \bibinfo{author}{\bibfnamefont{W.~T.} \bibnamefont{Giele}},
  \bibinfo{author}{\bibfnamefont{X.}~\bibnamefont{Liu}}, \bibnamefont{and}
  \bibinfo{author}{\bibfnamefont{F.}~\bibnamefont{Petriello}},
  \bibinfo{journal}{Phys. Rev. Lett.} \textbf{\bibinfo{volume}{116}},
  \bibinfo{pages}{152001} (\bibinfo{year}{2016}{\natexlab{a}}),
  \eprint{1512.01291}.

\bibitem[{\citenamefont{Boughezal
  et~al.}(2016{\natexlab{b}})\citenamefont{Boughezal, Liu, and
  Petriello}}]{Boughezal:2016dtm}
\bibinfo{author}{\bibfnamefont{R.}~\bibnamefont{Boughezal}},
  \bibinfo{author}{\bibfnamefont{X.}~\bibnamefont{Liu}}, \bibnamefont{and}
  \bibinfo{author}{\bibfnamefont{F.}~\bibnamefont{Petriello}},
  \bibinfo{journal}{Phys. Rev.} \textbf{\bibinfo{volume}{D94}},
  \bibinfo{pages}{113009} (\bibinfo{year}{2016}{\natexlab{b}}),
  \eprint{1602.06965}.

\bibitem[{\citenamefont{Gehrmann-De~Ridder
  et~al.}(2016{\natexlab{b}})\citenamefont{Gehrmann-De~Ridder, Gehrmann,
  Glover, Huss, and Morgan}}]{Ridder:2016nkl}
\bibinfo{author}{\bibfnamefont{A.}~\bibnamefont{Gehrmann-De~Ridder}},
  \bibinfo{author}{\bibfnamefont{T.}~\bibnamefont{Gehrmann}},
  \bibinfo{author}{\bibfnamefont{E.~W.~N.} \bibnamefont{Glover}},
  \bibinfo{author}{\bibfnamefont{A.}~\bibnamefont{Huss}}, \bibnamefont{and}
  \bibinfo{author}{\bibfnamefont{T.~A.} \bibnamefont{Morgan}},
  \bibinfo{journal}{JHEP} \textbf{\bibinfo{volume}{07}}, \bibinfo{pages}{133}
  (\bibinfo{year}{2016}{\natexlab{b}}), \eprint{1605.04295}.

\bibitem[{\citenamefont{Gehrmann-De~Ridder
  et~al.}(2018)\citenamefont{Gehrmann-De~Ridder, Gehrmann, Glover, Huss, and
  Walker}}]{Gehrmann-DeRidder:2017mvr}
\bibinfo{author}{\bibfnamefont{A.}~\bibnamefont{Gehrmann-De~Ridder}},
  \bibinfo{author}{\bibfnamefont{T.}~\bibnamefont{Gehrmann}},
  \bibinfo{author}{\bibfnamefont{E.~W.~N.} \bibnamefont{Glover}},
  \bibinfo{author}{\bibfnamefont{A.}~\bibnamefont{Huss}}, \bibnamefont{and}
  \bibinfo{author}{\bibfnamefont{D.~M.} \bibnamefont{Walker}},
  \bibinfo{journal}{Phys. Rev. Lett.} \textbf{\bibinfo{volume}{120}},
  \bibinfo{pages}{122001} (\bibinfo{year}{2018}), \eprint{1712.07543}.

\bibitem[{\citenamefont{Giele et~al.}(1993)\citenamefont{Giele, Glover, and
  Kosower}}]{Giele:1993dj}
\bibinfo{author}{\bibfnamefont{W.}~\bibnamefont{Giele}},
  \bibinfo{author}{\bibfnamefont{E.~N.} \bibnamefont{Glover}},
  \bibnamefont{and} \bibinfo{author}{\bibfnamefont{D.~A.}
  \bibnamefont{Kosower}}, \bibinfo{journal}{Nucl.Phys.}
  \textbf{\bibinfo{volume}{B403}}, \bibinfo{pages}{633} (\bibinfo{year}{1993}),
  \eprint{hep-ph/9302225}.

\bibitem[{\citenamefont{Campbell and Ellis}(2002)}]{Campbell:2002tg}
\bibinfo{author}{\bibfnamefont{J.~M.} \bibnamefont{Campbell}} \bibnamefont{and}
  \bibinfo{author}{\bibfnamefont{R.}~\bibnamefont{Ellis}},
  \bibinfo{journal}{Phys. Rev.} \textbf{\bibinfo{volume}{D65}},
  \bibinfo{pages}{113007} (\bibinfo{year}{2002}), \eprint{hep-ph/0202176}.

\bibitem[{\citenamefont{Campbell et~al.}(2003)\citenamefont{Campbell, Ellis,
  and Rainwater}}]{Campbell:2003hd}
\bibinfo{author}{\bibfnamefont{J.~M.} \bibnamefont{Campbell}},
  \bibinfo{author}{\bibfnamefont{R.}~\bibnamefont{Ellis}}, \bibnamefont{and}
  \bibinfo{author}{\bibfnamefont{D.~L.} \bibnamefont{Rainwater}},
  \bibinfo{journal}{Phys. Rev.} \textbf{\bibinfo{volume}{D68}},
  \bibinfo{pages}{094021} (\bibinfo{year}{2003}), \eprint{hep-ph/0308195}.

\bibitem[{\citenamefont{Ellis et~al.}(2009)\citenamefont{Ellis, Melnikov, and
  Zanderighi}}]{Ellis:2009zw}
\bibinfo{author}{\bibfnamefont{R.}~\bibnamefont{Ellis}},
  \bibinfo{author}{\bibfnamefont{K.}~\bibnamefont{Melnikov}}, \bibnamefont{and}
  \bibinfo{author}{\bibfnamefont{G.}~\bibnamefont{Zanderighi}},
  \bibinfo{journal}{JHEP} \textbf{\bibinfo{volume}{04}}, \bibinfo{pages}{077}
  (\bibinfo{year}{2009}), \eprint{0901.4101}.

\bibitem[{\citenamefont{Berger et~al.}(2009)\citenamefont{Berger, Bern, Dixon,
  Febres-Cordero, Forde, Gleisberg, Ita, Kosower, and
  Ma{\^i}tre}}]{Berger:2009zg}
\bibinfo{author}{\bibfnamefont{C.~F.} \bibnamefont{Berger}},
  \bibinfo{author}{\bibfnamefont{Z.}~\bibnamefont{Bern}},
  \bibinfo{author}{\bibfnamefont{L.~J.} \bibnamefont{Dixon}},
  \bibinfo{author}{\bibfnamefont{F.}~\bibnamefont{Febres-Cordero}},
  \bibinfo{author}{\bibfnamefont{D.}~\bibnamefont{Forde}},
  \bibinfo{author}{\bibfnamefont{T.}~\bibnamefont{Gleisberg}},
  \bibinfo{author}{\bibfnamefont{H.}~\bibnamefont{Ita}},
  \bibinfo{author}{\bibfnamefont{D.~A.} \bibnamefont{Kosower}},
  \bibnamefont{and}
  \bibinfo{author}{\bibfnamefont{D.}~\bibnamefont{Ma{\^i}tre}},
  \bibinfo{journal}{Phys. Rev. Lett.} \textbf{\bibinfo{volume}{102}},
  \bibinfo{pages}{222001} (\bibinfo{year}{2009}), \eprint{0902.2760}.

\bibitem[{\citenamefont{Berger et~al.}(2010)\citenamefont{Berger, Bern, Dixon,
  Febres-Cordero, Forde, Gleisberg, Ita, Kosower, and
  Ma{\^i}tre}}]{Berger:2010vm}
\bibinfo{author}{\bibfnamefont{C.~F.} \bibnamefont{Berger}},
  \bibinfo{author}{\bibfnamefont{Z.}~\bibnamefont{Bern}},
  \bibinfo{author}{\bibfnamefont{L.~J.} \bibnamefont{Dixon}},
  \bibinfo{author}{\bibfnamefont{F.}~\bibnamefont{Febres-Cordero}},
  \bibinfo{author}{\bibfnamefont{D.}~\bibnamefont{Forde}},
  \bibinfo{author}{\bibfnamefont{T.}~\bibnamefont{Gleisberg}},
  \bibinfo{author}{\bibfnamefont{H.}~\bibnamefont{Ita}},
  \bibinfo{author}{\bibfnamefont{D.~A.} \bibnamefont{Kosower}},
  \bibnamefont{and}
  \bibinfo{author}{\bibfnamefont{D.}~\bibnamefont{Ma{\^i}tre}},
  \bibinfo{journal}{Phys. Rev.} \textbf{\bibinfo{volume}{D82}},
  \bibinfo{pages}{074002} (\bibinfo{year}{2010}), \eprint{1004.1659}.

\bibitem[{\citenamefont{Berger et~al.}(2011)\citenamefont{Berger, Bern, Dixon,
  Febres-Cordero, Forde, Gleisberg, Ita, Kosower, and
  Ma{\^i}tre}}]{Berger:2010zx}
\bibinfo{author}{\bibfnamefont{C.~F.} \bibnamefont{Berger}},
  \bibinfo{author}{\bibfnamefont{Z.}~\bibnamefont{Bern}},
  \bibinfo{author}{\bibfnamefont{L.~J.} \bibnamefont{Dixon}},
  \bibinfo{author}{\bibfnamefont{F.}~\bibnamefont{Febres-Cordero}},
  \bibinfo{author}{\bibfnamefont{D.}~\bibnamefont{Forde}},
  \bibinfo{author}{\bibfnamefont{T.}~\bibnamefont{Gleisberg}},
  \bibinfo{author}{\bibfnamefont{H.}~\bibnamefont{Ita}},
  \bibinfo{author}{\bibfnamefont{D.~A.} \bibnamefont{Kosower}},
  \bibnamefont{and}
  \bibinfo{author}{\bibfnamefont{D.}~\bibnamefont{Ma{\^i}tre}},
  \bibinfo{journal}{Phys. Rev. Lett.} \textbf{\bibinfo{volume}{106}},
  \bibinfo{pages}{092001} (\bibinfo{year}{2011}), \eprint{1009.2338}.

\bibitem[{\citenamefont{Ita et~al.}(2012)\citenamefont{Ita, Bern, Dixon,
  Febres-Cordero, Kosower, and Ma{\^i}tre}}]{Ita:2011wn}
\bibinfo{author}{\bibfnamefont{H.}~\bibnamefont{Ita}},
  \bibinfo{author}{\bibfnamefont{Z.}~\bibnamefont{Bern}},
  \bibinfo{author}{\bibfnamefont{L.~J.} \bibnamefont{Dixon}},
  \bibinfo{author}{\bibfnamefont{F.}~\bibnamefont{Febres-Cordero}},
  \bibinfo{author}{\bibfnamefont{D.~A.} \bibnamefont{Kosower}},
  \bibnamefont{and}
  \bibinfo{author}{\bibfnamefont{D.}~\bibnamefont{Ma{\^i}tre}},
  \bibinfo{journal}{Phys.Rev.} \textbf{\bibinfo{volume}{D85}},
  \bibinfo{pages}{031501} (\bibinfo{year}{2012}), \eprint{1108.2229}.

\bibitem[{\citenamefont{Bern et~al.}(2013)\citenamefont{Bern, Dixon,
  Febres~Cordero, H{\"o}che, Ita, Kosower, Ma{\^i}tre, and
  Ozeren}}]{Bern:2013gka}
\bibinfo{author}{\bibfnamefont{Z.}~\bibnamefont{Bern}},
  \bibinfo{author}{\bibfnamefont{L.}~\bibnamefont{Dixon}},
  \bibinfo{author}{\bibfnamefont{F.}~\bibnamefont{Febres~Cordero}},
  \bibinfo{author}{\bibfnamefont{S.}~\bibnamefont{H{\"o}che}},
  \bibinfo{author}{\bibfnamefont{H.}~\bibnamefont{Ita}},
  \bibinfo{author}{\bibfnamefont{D.~A.} \bibnamefont{Kosower}},
  \bibinfo{author}{\bibfnamefont{D.}~\bibnamefont{Ma{\^i}tre}},
  \bibnamefont{and} \bibinfo{author}{\bibfnamefont{K.~J.}
  \bibnamefont{Ozeren}}, \bibinfo{journal}{Phys.Rev.}
  \textbf{\bibinfo{volume}{D88}}, \bibinfo{pages}{014025}
  (\bibinfo{year}{2013}), \eprint{1304.1253}.

\bibitem[{\citenamefont{Kanaki and Papadopoulos}(2000)}]{Kanaki:2000ey}
\bibinfo{author}{\bibfnamefont{A.}~\bibnamefont{Kanaki}} \bibnamefont{and}
  \bibinfo{author}{\bibfnamefont{C.~G.} \bibnamefont{Papadopoulos}},
  \bibinfo{journal}{Comput. Phys. Commun.} \textbf{\bibinfo{volume}{132}},
  \bibinfo{pages}{306} (\bibinfo{year}{2000}), \eprint{hep-ph/0002082}.

\bibitem[{\citenamefont{Papadopoulos}(2001)}]{Papadopoulos:2000tt}
\bibinfo{author}{\bibfnamefont{C.~G.} \bibnamefont{Papadopoulos}},
  \bibinfo{journal}{Comput. Phys. Commun.} \textbf{\bibinfo{volume}{137}},
  \bibinfo{pages}{247} (\bibinfo{year}{2001}), \eprint{hep-ph/0007335}.

\bibitem[{\citenamefont{Krauss et~al.}(2002)\citenamefont{Krauss, Kuhn, and
  Soff}}]{Krauss:2001iv}
\bibinfo{author}{\bibfnamefont{F.}~\bibnamefont{Krauss}},
  \bibinfo{author}{\bibfnamefont{R.}~\bibnamefont{Kuhn}}, \bibnamefont{and}
  \bibinfo{author}{\bibfnamefont{G.}~\bibnamefont{Soff}},
  \bibinfo{journal}{JHEP} \textbf{\bibinfo{volume}{02}}, \bibinfo{pages}{044}
  (\bibinfo{year}{2002}), \eprint{hep-ph/0109036}.

\bibitem[{\citenamefont{Mangano et~al.}(2003)\citenamefont{Mangano, Moretti,
  Piccinini, Pittau, and Polosa}}]{Mangano:2002ea}
\bibinfo{author}{\bibfnamefont{M.~L.} \bibnamefont{Mangano}},
  \bibinfo{author}{\bibfnamefont{M.}~\bibnamefont{Moretti}},
  \bibinfo{author}{\bibfnamefont{F.}~\bibnamefont{Piccinini}},
  \bibinfo{author}{\bibfnamefont{R.}~\bibnamefont{Pittau}}, \bibnamefont{and}
  \bibinfo{author}{\bibfnamefont{A.~D.} \bibnamefont{Polosa}},
  \bibinfo{journal}{JHEP} \textbf{\bibinfo{volume}{07}}, \bibinfo{pages}{001}
  (\bibinfo{year}{2003}), \eprint{hep-ph/0206293}.

\bibitem[{\citenamefont{Gleisberg and H{\"o}che}(2008)}]{Gleisberg:2008fv}
\bibinfo{author}{\bibfnamefont{T.}~\bibnamefont{Gleisberg}} \bibnamefont{and}
  \bibinfo{author}{\bibfnamefont{S.}~\bibnamefont{H{\"o}che}},
  \bibinfo{journal}{JHEP} \textbf{\bibinfo{volume}{12}}, \bibinfo{pages}{039}
  (\bibinfo{year}{2008}), \eprint{0808.3674}.

\bibitem[{\citenamefont{Alwall et~al.}(2011)\citenamefont{Alwall, Herquet,
  Maltoni, Mattelaer, and Stelzer}}]{Alwall:2011uj}
\bibinfo{author}{\bibfnamefont{J.}~\bibnamefont{Alwall}},
  \bibinfo{author}{\bibfnamefont{M.}~\bibnamefont{Herquet}},
  \bibinfo{author}{\bibfnamefont{F.}~\bibnamefont{Maltoni}},
  \bibinfo{author}{\bibfnamefont{O.}~\bibnamefont{Mattelaer}},
  \bibnamefont{and} \bibinfo{author}{\bibfnamefont{T.}~\bibnamefont{Stelzer}},
  \bibinfo{journal}{JHEP} \textbf{\bibinfo{volume}{06}}, \bibinfo{pages}{128}
  (\bibinfo{year}{2011}), \eprint{1106.0522}.

\bibitem[{\citenamefont{Alwall et~al.}(2014)\citenamefont{Alwall, Frederix,
  Frixione, Hirschi, Maltoni, Mattelaer, Shao, Stelzer, Torrielli, and
  Zaro}}]{Alwall:2014hca}
\bibinfo{author}{\bibfnamefont{J.}~\bibnamefont{Alwall}},
  \bibinfo{author}{\bibfnamefont{R.}~\bibnamefont{Frederix}},
  \bibinfo{author}{\bibfnamefont{S.}~\bibnamefont{Frixione}},
  \bibinfo{author}{\bibfnamefont{V.}~\bibnamefont{Hirschi}},
  \bibinfo{author}{\bibfnamefont{F.}~\bibnamefont{Maltoni}},
  \bibinfo{author}{\bibfnamefont{O.}~\bibnamefont{Mattelaer}},
  \bibinfo{author}{\bibfnamefont{H.-S.} \bibnamefont{Shao}},
  \bibinfo{author}{\bibfnamefont{T.}~\bibnamefont{Stelzer}},
  \bibinfo{author}{\bibfnamefont{P.}~\bibnamefont{Torrielli}},
  \bibnamefont{and} \bibinfo{author}{\bibfnamefont{M.}~\bibnamefont{Zaro}},
  \bibinfo{journal}{JHEP} \textbf{\bibinfo{volume}{07}}, \bibinfo{pages}{079}
  (\bibinfo{year}{2014}), \eprint{1405.0301}.

\bibitem[{\citenamefont{Andr\'e and Sj{\"o}strand}(1998)}]{Andre:1997vh}
\bibinfo{author}{\bibfnamefont{J.}~\bibnamefont{Andr\'e}} \bibnamefont{and}
  \bibinfo{author}{\bibfnamefont{T.}~\bibnamefont{Sj{\"o}strand}},
  \bibinfo{journal}{Phys. Rev.} \textbf{\bibinfo{volume}{D57}},
  \bibinfo{pages}{5767} (\bibinfo{year}{1998}), \eprint{hep-ph/9708390}.

\bibitem[{\citenamefont{Mangano et~al.}(2002)\citenamefont{Mangano, Moretti,
  and Pittau}}]{Mangano:2001xp}
\bibinfo{author}{\bibfnamefont{M.~L.} \bibnamefont{Mangano}},
  \bibinfo{author}{\bibfnamefont{M.}~\bibnamefont{Moretti}}, \bibnamefont{and}
  \bibinfo{author}{\bibfnamefont{R.}~\bibnamefont{Pittau}},
  \bibinfo{journal}{Nucl. Phys.} \textbf{\bibinfo{volume}{B632}},
  \bibinfo{pages}{343} (\bibinfo{year}{2002}), \eprint{hep-ph/0108069}.

\bibitem[{\citenamefont{Catani et~al.}(2001)\citenamefont{Catani, Krauss, Kuhn,
  and Webber}}]{Catani:2001cc}
\bibinfo{author}{\bibfnamefont{S.}~\bibnamefont{Catani}},
  \bibinfo{author}{\bibfnamefont{F.}~\bibnamefont{Krauss}},
  \bibinfo{author}{\bibfnamefont{R.}~\bibnamefont{Kuhn}}, \bibnamefont{and}
  \bibinfo{author}{\bibfnamefont{B.~R.} \bibnamefont{Webber}},
  \bibinfo{journal}{JHEP} \textbf{\bibinfo{volume}{11}}, \bibinfo{pages}{063}
  (\bibinfo{year}{2001}), \eprint{hep-ph/0109231}.

\bibitem[{\citenamefont{L{\"o}nnblad}(2002)}]{Lonnblad:2001iq}
\bibinfo{author}{\bibfnamefont{L.}~\bibnamefont{L{\"o}nnblad}},
  \bibinfo{journal}{JHEP} \textbf{\bibinfo{volume}{05}}, \bibinfo{pages}{046}
  (\bibinfo{year}{2002}), \eprint{hep-ph/0112284}.

\bibitem[{\citenamefont{Hamilton et~al.}(2009)\citenamefont{Hamilton,
  Richardson, and Tully}}]{Hamilton:2009ne}
\bibinfo{author}{\bibfnamefont{K.}~\bibnamefont{Hamilton}},
  \bibinfo{author}{\bibfnamefont{P.}~\bibnamefont{Richardson}},
  \bibnamefont{and} \bibinfo{author}{\bibfnamefont{J.}~\bibnamefont{Tully}},
  \bibinfo{journal}{JHEP} \textbf{\bibinfo{volume}{11}}, \bibinfo{pages}{038}
  (\bibinfo{year}{2009}), \eprint{0905.3072}.

\bibitem[{\citenamefont{H{\"o}che et~al.}(2011)\citenamefont{H{\"o}che, Krauss,
  Sch{\"o}nherr, and Siegert}}]{Hoeche:2010kg}
\bibinfo{author}{\bibfnamefont{S.}~\bibnamefont{H{\"o}che}},
  \bibinfo{author}{\bibfnamefont{F.}~\bibnamefont{Krauss}},
  \bibinfo{author}{\bibfnamefont{M.}~\bibnamefont{Sch{\"o}nherr}},
  \bibnamefont{and} \bibinfo{author}{\bibfnamefont{F.}~\bibnamefont{Siegert}},
  \bibinfo{journal}{JHEP} \textbf{\bibinfo{volume}{08}}, \bibinfo{pages}{123}
  (\bibinfo{year}{2011}), \eprint{1009.1127}.

\bibitem[{\citenamefont{L{\"o}nnblad and Prestel}(2013)}]{Lonnblad:2012ng}
\bibinfo{author}{\bibfnamefont{L.}~\bibnamefont{L{\"o}nnblad}}
  \bibnamefont{and} \bibinfo{author}{\bibfnamefont{S.}~\bibnamefont{Prestel}},
  \bibinfo{journal}{JHEP} \textbf{\bibinfo{volume}{02}}, \bibinfo{pages}{094}
  (\bibinfo{year}{2013}), \eprint{1211.4827}.

\bibitem[{\citenamefont{Pl{\"a}tzer}(2013)}]{Platzer:2012bs}
\bibinfo{author}{\bibfnamefont{S.}~\bibnamefont{Pl{\"a}tzer}},
  \bibinfo{journal}{JHEP} \textbf{\bibinfo{volume}{08}}, \bibinfo{pages}{114}
  (\bibinfo{year}{2013}), \eprint{1211.5467}.

\bibitem[{\citenamefont{Buckley et~al.}(2011)}]{Buckley:2011ms}
\bibinfo{author}{\bibfnamefont{A.}~\bibnamefont{Buckley}} \bibnamefont{et~al.},
  \bibinfo{journal}{Phys. Rept.} \textbf{\bibinfo{volume}{504}},
  \bibinfo{pages}{145} (\bibinfo{year}{2011}), \eprint{1101.2599}.

\bibitem[{\citenamefont{Berends and Giele}(1987)}]{Berends:1987cv}
\bibinfo{author}{\bibfnamefont{F.~A.} \bibnamefont{Berends}} \bibnamefont{and}
  \bibinfo{author}{\bibfnamefont{W.}~\bibnamefont{Giele}},
  \bibinfo{journal}{Nucl. Phys.} \textbf{\bibinfo{volume}{B294}},
  \bibinfo{pages}{700} (\bibinfo{year}{1987}).

\bibitem[{\citenamefont{Berends and Giele}(1988)}]{Berends:1987me}
\bibinfo{author}{\bibfnamefont{F.~A.} \bibnamefont{Berends}} \bibnamefont{and}
  \bibinfo{author}{\bibfnamefont{W.~T.} \bibnamefont{Giele}},
  \bibinfo{journal}{Nucl. Phys.} \textbf{\bibinfo{volume}{B306}},
  \bibinfo{pages}{759} (\bibinfo{year}{1988}).

\bibitem[{\citenamefont{Ballestrero and Maina}(1995)}]{Ballestrero:1994jn}
\bibinfo{author}{\bibfnamefont{A.}~\bibnamefont{Ballestrero}} \bibnamefont{and}
  \bibinfo{author}{\bibfnamefont{E.}~\bibnamefont{Maina}},
  \bibinfo{journal}{Phys. Lett.} \textbf{\bibinfo{volume}{B350}},
  \bibinfo{pages}{225} (\bibinfo{year}{1995}), \eprint{hep-ph/9403244}.

\bibitem[{\citenamefont{Aad et~al.}(2013)}]{Aad:2013ysa}
\bibinfo{author}{\bibfnamefont{G.}~\bibnamefont{Aad}} \bibnamefont{et~al.}
  (\bibinfo{collaboration}{ATLAS}), \bibinfo{journal}{JHEP}
  \textbf{\bibinfo{volume}{07}}, \bibinfo{pages}{032} (\bibinfo{year}{2013}),
  \eprint{1304.7098}.

\bibitem[{\citenamefont{Aad et~al.}(2015)}]{Aad:2014qxa}
\bibinfo{author}{\bibfnamefont{G.}~\bibnamefont{Aad}} \bibnamefont{et~al.}
  (\bibinfo{collaboration}{ATLAS}), \bibinfo{journal}{Eur. Phys. J.}
  \textbf{\bibinfo{volume}{C75}}, \bibinfo{pages}{82} (\bibinfo{year}{2015}),
  \eprint{1409.8639}.

\bibitem[{\citenamefont{Aaboud et~al.}(2018)}]{Aaboud:2017soa}
\bibinfo{author}{\bibfnamefont{M.}~\bibnamefont{Aaboud}} \bibnamefont{et~al.}
  (\bibinfo{collaboration}{ATLAS}), \bibinfo{journal}{JHEP}
  \textbf{\bibinfo{volume}{05}}, \bibinfo{pages}{077} (\bibinfo{year}{2018}),
  \eprint{1711.03296}.

\bibitem[{\citenamefont{Aaboud et~al.}(2017)}]{Aaboud:2017hbk}
\bibinfo{author}{\bibfnamefont{M.}~\bibnamefont{Aaboud}} \bibnamefont{et~al.}
  (\bibinfo{collaboration}{ATLAS}), \bibinfo{journal}{Eur. Phys. J.}
  \textbf{\bibinfo{volume}{C77}}, \bibinfo{pages}{361} (\bibinfo{year}{2017}),
  \eprint{1702.05725}.

\bibitem[{\citenamefont{Khachatryan
  et~al.}(2017{\natexlab{a}})}]{Khachatryan:2016fue}
\bibinfo{author}{\bibfnamefont{V.}~\bibnamefont{Khachatryan}}
  \bibnamefont{et~al.} (\bibinfo{collaboration}{CMS}), \bibinfo{journal}{Phys.
  Rev.} \textbf{\bibinfo{volume}{D95}}, \bibinfo{pages}{052002}
  (\bibinfo{year}{2017}{\natexlab{a}}), \eprint{1610.04222}.

\bibitem[{\citenamefont{Sirunyan et~al.}(2017)}]{Sirunyan:2017wgx}
\bibinfo{author}{\bibfnamefont{A.~M.} \bibnamefont{Sirunyan}}
  \bibnamefont{et~al.} (\bibinfo{collaboration}{CMS}), \bibinfo{journal}{Phys.
  Rev.} \textbf{\bibinfo{volume}{D96}}, \bibinfo{pages}{072005}
  (\bibinfo{year}{2017}), \eprint{1707.05979}.

\bibitem[{\citenamefont{Sirunyan et~al.}(2018)}]{Sirunyan:2018cpw}
\bibinfo{author}{\bibfnamefont{A.~M.} \bibnamefont{Sirunyan}}
  \bibnamefont{et~al.} (\bibinfo{collaboration}{CMS}), \bibinfo{journal}{Eur.
  Phys. J.} \textbf{\bibinfo{volume}{C78}}, \bibinfo{pages}{965}
  (\bibinfo{year}{2018}), \eprint{1804.05252}.

\bibitem[{\citenamefont{Khachatryan
  et~al.}(2017{\natexlab{b}})}]{Khachatryan:2016crw}
\bibinfo{author}{\bibfnamefont{V.}~\bibnamefont{Khachatryan}}
  \bibnamefont{et~al.} (\bibinfo{collaboration}{CMS}), \bibinfo{journal}{JHEP}
  \textbf{\bibinfo{volume}{04}}, \bibinfo{pages}{022}
  (\bibinfo{year}{2017}{\natexlab{b}}), \eprint{1611.03844}.

\bibitem[{\citenamefont{Sj{\"o}strand et~al.}(2014)\citenamefont{Sj{\"o}strand,
  Ask, Christiansen, Corke, Desai, Ilten, Mrenna, Prestel, Rasmussen, and
  Skands}}]{Sjostrand:2014zea}
\bibinfo{author}{\bibfnamefont{T.}~\bibnamefont{Sj{\"o}strand}},
  \bibinfo{author}{\bibfnamefont{S.}~\bibnamefont{Ask}},
  \bibinfo{author}{\bibfnamefont{J.~R.} \bibnamefont{Christiansen}},
  \bibinfo{author}{\bibfnamefont{R.}~\bibnamefont{Corke}},
  \bibinfo{author}{\bibfnamefont{N.}~\bibnamefont{Desai}},
  \bibinfo{author}{\bibfnamefont{P.}~\bibnamefont{Ilten}},
  \bibinfo{author}{\bibfnamefont{S.}~\bibnamefont{Mrenna}},
  \bibinfo{author}{\bibfnamefont{S.}~\bibnamefont{Prestel}},
  \bibinfo{author}{\bibfnamefont{C.~O.} \bibnamefont{Rasmussen}},
  \bibnamefont{and} \bibinfo{author}{\bibfnamefont{P.~Z.} \bibnamefont{Skands}}
  (\bibinfo{year}{2014}), \eprint{1410.3012}.

\bibitem[{\citenamefont{H{\"o}che et~al.}(2013)\citenamefont{H{\"o}che, Reina,
  Wobisch et~al.}}]{Hoche:2013zja}
\bibinfo{author}{\bibfnamefont{S.}~\bibnamefont{H{\"o}che}},
  \bibinfo{author}{\bibfnamefont{L.}~\bibnamefont{Reina}},
  \bibinfo{author}{\bibfnamefont{M.}~\bibnamefont{Wobisch}},
  \bibnamefont{et~al.} (\bibinfo{year}{2013}), \eprint{1309.3598}.

\bibitem[{\citenamefont{Bauerdick et~al.}(2014)\citenamefont{Bauerdick,
  Gottlieb et~al.}}]{Bauerdick:2014qka}
\bibinfo{author}{\bibfnamefont{L.}~\bibnamefont{Bauerdick}},
  \bibinfo{author}{\bibfnamefont{S.}~\bibnamefont{Gottlieb}},
  \bibnamefont{et~al.} (\bibinfo{year}{2014}), \eprint{1401.6117}.

\bibitem[{\citenamefont{Childers et~al.}(2017)\citenamefont{Childers, Uram,
  LeCompte, Papka, and Benjamin}}]{Childers:2015tyv}
\bibinfo{author}{\bibfnamefont{J.~T.} \bibnamefont{Childers}},
  \bibinfo{author}{\bibfnamefont{T.~D.} \bibnamefont{Uram}},
  \bibinfo{author}{\bibfnamefont{T.~J.} \bibnamefont{LeCompte}},
  \bibinfo{author}{\bibfnamefont{M.~E.} \bibnamefont{Papka}}, \bibnamefont{and}
  \bibinfo{author}{\bibfnamefont{D.~P.} \bibnamefont{Benjamin}},
  \bibinfo{journal}{Comput. Phys. Commun.} \textbf{\bibinfo{volume}{210}},
  \bibinfo{pages}{54} (\bibinfo{year}{2017}), \eprint{1511.07312}.

\bibitem[{\citenamefont{Benjamin et~al.}(2017)\citenamefont{Benjamin, Childers,
  Hoeche, LeCompte, and Uram}}]{Benjamin:2017xdd}
\bibinfo{author}{\bibfnamefont{D.}~\bibnamefont{Benjamin}},
  \bibinfo{author}{\bibfnamefont{J.~T.} \bibnamefont{Childers}},
  \bibinfo{author}{\bibfnamefont{S.}~\bibnamefont{Hoeche}},
  \bibinfo{author}{\bibfnamefont{T.}~\bibnamefont{LeCompte}}, \bibnamefont{and}
  \bibinfo{author}{\bibfnamefont{T.}~\bibnamefont{Uram}}, \bibinfo{journal}{J.
  Phys. Conf. Ser.} \textbf{\bibinfo{volume}{898}}, \bibinfo{pages}{072044}
  (\bibinfo{year}{2017}).

\bibitem[{\citenamefont{Schulz et~al.}(2019{\natexlab{a}})\citenamefont{Schulz,
  H{\"o}che, and Prestel}}]{schulz_holger_2019_2678039}
\bibinfo{author}{\bibfnamefont{H.}~\bibnamefont{Schulz}},
  \bibinfo{author}{\bibfnamefont{S.}~\bibnamefont{H{\"o}che}},
  \bibnamefont{and} \bibinfo{author}{\bibfnamefont{S.}~\bibnamefont{Prestel}},
  \emph{\bibinfo{title}{{$Z$ + up to 9 jets parton level events at 14 TeV in
  HDF5}}} (\bibinfo{year}{2019}{\natexlab{a}}),
  \urlprefix\url{https://doi.org/10.5281/zenodo.2678039}.

\bibitem[{\citenamefont{Schulz et~al.}(2019{\natexlab{b}})\citenamefont{Schulz,
  H{\"o}che, and Prestel}}]{schulz_holger_2019_2678055}
\bibinfo{author}{\bibfnamefont{H.}~\bibnamefont{Schulz}},
  \bibinfo{author}{\bibfnamefont{S.}~\bibnamefont{H{\"o}che}},
  \bibnamefont{and} \bibinfo{author}{\bibfnamefont{S.}~\bibnamefont{Prestel}},
  \emph{\bibinfo{title}{{$W^+$ + up to 9 jets parton level events at 14 TeV in
  HDF5}}} (\bibinfo{year}{2019}{\natexlab{b}}),
  \urlprefix\url{https://doi.org/10.5281/zenodo.2678055}.

\bibitem[{\citenamefont{Schulz et~al.}(2019{\natexlab{c}})\citenamefont{Schulz,
  H{\"o}che, and Prestel}}]{schulz_holger_2019_2678091}
\bibinfo{author}{\bibfnamefont{H.}~\bibnamefont{Schulz}},
  \bibinfo{author}{\bibfnamefont{S.}~\bibnamefont{H{\"o}che}},
  \bibnamefont{and} \bibinfo{author}{\bibfnamefont{S.}~\bibnamefont{Prestel}},
  \emph{\bibinfo{title}{{$W^-$ + up to 9 jets parton level events at 14 TeV in
  HDF5}}} (\bibinfo{year}{2019}{\natexlab{c}}),
  \urlprefix\url{https://doi.org/10.5281/zenodo.2678091}.

\bibitem[{\citenamefont{H{\"o}che et~al.}(2019)\citenamefont{H{\"o}che,
  Prestel, and Schulz}}]{hpcgenrepo}
\bibinfo{author}{\bibfnamefont{S.}~\bibnamefont{H{\"o}che}},
  \bibinfo{author}{\bibfnamefont{S.}~\bibnamefont{Prestel}}, \bibnamefont{and}
  \bibinfo{author}{\bibfnamefont{H.}~\bibnamefont{Schulz}}
  (\bibinfo{year}{2019}), \urlprefix\url{https://gitlab.com/hpcgen}.

\bibitem[{\citenamefont{Amati et~al.}(1980)\citenamefont{Amati, Bassetto,
  Ciafaloni, Marchesini, and Veneziano}}]{Amati:1980ch}
\bibinfo{author}{\bibfnamefont{D.}~\bibnamefont{Amati}},
  \bibinfo{author}{\bibfnamefont{A.}~\bibnamefont{Bassetto}},
  \bibinfo{author}{\bibfnamefont{M.}~\bibnamefont{Ciafaloni}},
  \bibinfo{author}{\bibfnamefont{G.}~\bibnamefont{Marchesini}},
  \bibnamefont{and}
  \bibinfo{author}{\bibfnamefont{G.}~\bibnamefont{Veneziano}},
  \bibinfo{journal}{Nucl. Phys.} \textbf{\bibinfo{volume}{B173}},
  \bibinfo{pages}{429} (\bibinfo{year}{1980}).

\bibitem[{\citenamefont{Catani et~al.}(1991)\citenamefont{Catani, Webber, and
  Marchesini}}]{Catani:1990rr}
\bibinfo{author}{\bibfnamefont{S.}~\bibnamefont{Catani}},
  \bibinfo{author}{\bibfnamefont{B.~R.} \bibnamefont{Webber}},
  \bibnamefont{and}
  \bibinfo{author}{\bibfnamefont{G.}~\bibnamefont{Marchesini}},
  \bibinfo{journal}{Nucl. Phys.} \textbf{\bibinfo{volume}{B349}},
  \bibinfo{pages}{635} (\bibinfo{year}{1991}).

\bibitem[{\citenamefont{Salam}(2010)}]{Salam:2009jx}
\bibinfo{author}{\bibfnamefont{G.~P.} \bibnamefont{Salam}},
  \bibinfo{journal}{Eur. Phys. J.} \textbf{\bibinfo{volume}{C67}},
  \bibinfo{pages}{637} (\bibinfo{year}{2010}), \eprint{0906.1833}.

\bibitem[{\citenamefont{Alwall et~al.}(2007)}]{Alwall:2006yp}
\bibinfo{author}{\bibfnamefont{J.}~\bibnamefont{Alwall}} \bibnamefont{et~al.},
  \bibinfo{journal}{Comput. Phys. Commun.} \textbf{\bibinfo{volume}{176}},
  \bibinfo{pages}{300} (\bibinfo{year}{2007}), \eprint{hep-ph/0609017}.

\bibitem[{\citenamefont{Lepage}(1978)}]{Lepage:1977sw}
\bibinfo{author}{\bibfnamefont{G.~P.} \bibnamefont{Lepage}},
  \bibinfo{journal}{J. Comput. Phys.} \textbf{\bibinfo{volume}{27}},
  \bibinfo{pages}{192} (\bibinfo{year}{1978}).

\bibitem[{\citenamefont{Kleiss and Pittau}(1994)}]{Kleiss:1994qy}
\bibinfo{author}{\bibfnamefont{R.}~\bibnamefont{Kleiss}} \bibnamefont{and}
  \bibinfo{author}{\bibfnamefont{R.}~\bibnamefont{Pittau}},
  \bibinfo{journal}{Comput. Phys. Commun.} \textbf{\bibinfo{volume}{83}},
  \bibinfo{pages}{141} (\bibinfo{year}{1994}), \eprint{hep-ph/9405257}.

\bibitem[{\citenamefont{Giele et~al.}(2011)\citenamefont{Giele, Stavenga, and
  Winter}}]{Giele:2010ks}
\bibinfo{author}{\bibfnamefont{W.}~\bibnamefont{Giele}},
  \bibinfo{author}{\bibfnamefont{G.}~\bibnamefont{Stavenga}}, \bibnamefont{and}
  \bibinfo{author}{\bibfnamefont{J.-C.} \bibnamefont{Winter}},
  \bibinfo{journal}{Eur. Phys. J.} \textbf{\bibinfo{volume}{C71}},
  \bibinfo{pages}{1703} (\bibinfo{year}{2011}), \eprint{1002.3446}.

\bibitem[{\citenamefont{Campbell et~al.}(2015)\citenamefont{Campbell, Ellis,
  and Giele}}]{Campbell:2015qma}
\bibinfo{author}{\bibfnamefont{J.~M.} \bibnamefont{Campbell}},
  \bibinfo{author}{\bibfnamefont{R.~K.} \bibnamefont{Ellis}}, \bibnamefont{and}
  \bibinfo{author}{\bibfnamefont{W.~T.} \bibnamefont{Giele}},
  \bibinfo{journal}{Eur. Phys. J.} \textbf{\bibinfo{volume}{C75}},
  \bibinfo{pages}{246} (\bibinfo{year}{2015}), \eprint{1503.06182}.

\bibitem[{\citenamefont{{The Libzippp Developers}}(2019)}]{libzippp}
\bibinfo{author}{\bibnamefont{{The Libzippp Developers}}}
  (\bibinfo{year}{2019}), \urlprefix\url{https://github.com/ctabin/libzippp}.

\bibitem[{\citenamefont{Buckley et~al.}(2015)\citenamefont{Buckley, Ferrando,
  Lloyd, Nordstr{\"o}m, Page, R{\"u}fenacht, Sch{\"o}nherr, and
  Watt}}]{Buckley:2014ana}
\bibinfo{author}{\bibfnamefont{A.}~\bibnamefont{Buckley}},
  \bibinfo{author}{\bibfnamefont{J.}~\bibnamefont{Ferrando}},
  \bibinfo{author}{\bibfnamefont{S.}~\bibnamefont{Lloyd}},
  \bibinfo{author}{\bibfnamefont{K.}~\bibnamefont{Nordstr{\"o}m}},
  \bibinfo{author}{\bibfnamefont{B.}~\bibnamefont{Page}},
  \bibinfo{author}{\bibfnamefont{M.}~\bibnamefont{R{\"u}fenacht}},
  \bibinfo{author}{\bibfnamefont{M.}~\bibnamefont{Sch{\"o}nherr}},
  \bibnamefont{and} \bibinfo{author}{\bibfnamefont{G.}~\bibnamefont{Watt}},
  \bibinfo{journal}{Eur. Phys. J.} \textbf{\bibinfo{volume}{C75}},
  \bibinfo{pages}{132} (\bibinfo{year}{2015}), \eprint{1412.7420}.

\bibitem[{\citenamefont{Peterka et~al.}(2011)\citenamefont{Peterka, Ross,
  Kendall, Gyulassy, Pascucci, Shen, Lee, and Chaudhuri}}]{peterka_ldav11}
\bibinfo{author}{\bibfnamefont{T.}~\bibnamefont{Peterka}},
  \bibinfo{author}{\bibfnamefont{R.}~\bibnamefont{Ross}},
  \bibinfo{author}{\bibfnamefont{W.}~\bibnamefont{Kendall}},
  \bibinfo{author}{\bibfnamefont{A.}~\bibnamefont{Gyulassy}},
  \bibinfo{author}{\bibfnamefont{V.}~\bibnamefont{Pascucci}},
  \bibinfo{author}{\bibfnamefont{H.-W.} \bibnamefont{Shen}},
  \bibinfo{author}{\bibfnamefont{T.-Y.} \bibnamefont{Lee}}, \bibnamefont{and}
  \bibinfo{author}{\bibfnamefont{A.}~\bibnamefont{Chaudhuri}}, in
  \emph{\bibinfo{booktitle}{Proceedings of Large Data Analysis and
  Visualization Symposium LDAV'11}} (\bibinfo{address}{Providence, RI},
  \bibinfo{year}{2011}).

\bibitem[{\citenamefont{Morozov and Peterka}(2016)}]{morozov_ldav16}
\bibinfo{author}{\bibfnamefont{D.}~\bibnamefont{Morozov}} \bibnamefont{and}
  \bibinfo{author}{\bibfnamefont{T.}~\bibnamefont{Peterka}}, in
  \emph{\bibinfo{booktitle}{Proceedings of the 2016 IEEE Large Data Analysis
  and Visualization Symposium LDAV'16}} (\bibinfo{address}{Baltimore, MD},
  \bibinfo{year}{2016}).

\bibitem[{\citenamefont{{The HDF Group}}(2017)}]{HDF5}
\bibinfo{author}{\bibnamefont{{The HDF Group}}} (\bibinfo{year}{2017}),
  \urlprefix\url{https://support.hdfgroup.org/HDF5}.

\bibitem[{\citenamefont{Dulat et~al.}(2016)}]{Dulat:2015mca}
\bibinfo{author}{\bibfnamefont{S.}~\bibnamefont{Dulat}} \bibnamefont{et~al.},
  \bibinfo{journal}{Phys. Rev.} \textbf{\bibinfo{volume}{D93}},
  \bibinfo{pages}{033006} (\bibinfo{year}{2016}), \eprint{1506.07443}.

\bibitem[{\citenamefont{Gleisberg et~al.}(2009)\citenamefont{Gleisberg,
  H{\"o}che, Krauss, Sch\"{o}nherr, Schumann, Siegert, and
  Winter}}]{Gleisberg:2008ta}
\bibinfo{author}{\bibfnamefont{T.}~\bibnamefont{Gleisberg}},
  \bibinfo{author}{\bibfnamefont{S.}~\bibnamefont{H{\"o}che}},
  \bibinfo{author}{\bibfnamefont{F.}~\bibnamefont{Krauss}},
  \bibinfo{author}{\bibfnamefont{M.}~\bibnamefont{Sch\"{o}nherr}},
  \bibinfo{author}{\bibfnamefont{S.}~\bibnamefont{Schumann}},
  \bibinfo{author}{\bibfnamefont{F.}~\bibnamefont{Siegert}}, \bibnamefont{and}
  \bibinfo{author}{\bibfnamefont{J.}~\bibnamefont{Winter}},
  \bibinfo{journal}{JHEP} \textbf{\bibinfo{volume}{02}}, \bibinfo{pages}{007}
  (\bibinfo{year}{2009}), \eprint{0811.4622}.

\bibitem[{\citenamefont{Bellm et~al.}(2019)}]{Bellm:2019yyh}
\bibinfo{author}{\bibfnamefont{J.}~\bibnamefont{Bellm}} \bibnamefont{et~al.}
  (\bibinfo{year}{2019}), \eprint{1903.12563}.

\bibitem[{\citenamefont{Hamilton et~al.}(2012)\citenamefont{Hamilton, Nason,
  and Zanderighi}}]{Hamilton:2012np}
\bibinfo{author}{\bibfnamefont{K.}~\bibnamefont{Hamilton}},
  \bibinfo{author}{\bibfnamefont{P.}~\bibnamefont{Nason}}, \bibnamefont{and}
  \bibinfo{author}{\bibfnamefont{G.}~\bibnamefont{Zanderighi}},
  \bibinfo{journal}{JHEP} \textbf{\bibinfo{volume}{10}}, \bibinfo{pages}{155}
  (\bibinfo{year}{2012}), \eprint{1206.3572}.

\bibitem[{\citenamefont{Anger et~al.}(2018)\citenamefont{Anger, Febres~Cordero,
  H{\"o}che, and Ma{\^i}tre}}]{Anger:2017nkq}
\bibinfo{author}{\bibfnamefont{F.~R.} \bibnamefont{Anger}},
  \bibinfo{author}{\bibfnamefont{F.}~\bibnamefont{Febres~Cordero}},
  \bibinfo{author}{\bibfnamefont{S.}~\bibnamefont{H{\"o}che}},
  \bibnamefont{and}
  \bibinfo{author}{\bibfnamefont{D.}~\bibnamefont{Ma{\^i}tre}},
  \bibinfo{journal}{Phys. Rev.} \textbf{\bibinfo{volume}{D97}},
  \bibinfo{pages}{096010} (\bibinfo{year}{2018}), \eprint{1712.08621}.

\bibitem[{\citenamefont{Maltoni et~al.}(2003)\citenamefont{Maltoni, Paul,
  Stelzer, and Willenbrock}}]{Maltoni:2002mq}
\bibinfo{author}{\bibfnamefont{F.}~\bibnamefont{Maltoni}},
  \bibinfo{author}{\bibfnamefont{K.}~\bibnamefont{Paul}},
  \bibinfo{author}{\bibfnamefont{T.}~\bibnamefont{Stelzer}}, \bibnamefont{and}
  \bibinfo{author}{\bibfnamefont{S.}~\bibnamefont{Willenbrock}},
  \bibinfo{journal}{Phys. Rev.} \textbf{\bibinfo{volume}{D67}},
  \bibinfo{pages}{014026} (\bibinfo{year}{2003}), \eprint{hep-ph/0209271}.

\bibitem[{\citenamefont{Duhr et~al.}(2006)\citenamefont{Duhr, H{\"o}che, and
  Maltoni}}]{Duhr:2006iq}
\bibinfo{author}{\bibfnamefont{C.}~\bibnamefont{Duhr}},
  \bibinfo{author}{\bibfnamefont{S.}~\bibnamefont{H{\"o}che}},
  \bibnamefont{and} \bibinfo{author}{\bibfnamefont{F.}~\bibnamefont{Maltoni}},
  \bibinfo{journal}{JHEP} \textbf{\bibinfo{volume}{08}}, \bibinfo{pages}{062}
  (\bibinfo{year}{2006}), \eprint{hep-ph/0607057}.

\bibitem[{\citenamefont{{NERSC}}(2016)}]{cori}
\bibinfo{author}{\bibnamefont{{NERSC}}} (\bibinfo{year}{2016}),
  \urlprefix\url{https://www.nersc.gov/users/computational-systems/cori}.

\bibitem[{\citenamefont{Catani et~al.}(1993)\citenamefont{Catani, Dokshitzer,
  Seymour, and Webber}}]{Catani:1993hr}
\bibinfo{author}{\bibfnamefont{S.}~\bibnamefont{Catani}},
  \bibinfo{author}{\bibfnamefont{Y.~L.} \bibnamefont{Dokshitzer}},
  \bibinfo{author}{\bibfnamefont{M.~H.} \bibnamefont{Seymour}},
  \bibnamefont{and} \bibinfo{author}{\bibfnamefont{B.~R.}
  \bibnamefont{Webber}}, \bibinfo{journal}{Nucl. Phys.}
  \textbf{\bibinfo{volume}{B406}}, \bibinfo{pages}{187} (\bibinfo{year}{1993}).

\bibitem[{\citenamefont{Krauss et~al.}(2004)\citenamefont{Krauss,
  Sch{\"a}licke, Schumann, and Soff}}]{Krauss:2004bs}
\bibinfo{author}{\bibfnamefont{F.}~\bibnamefont{Krauss}},
  \bibinfo{author}{\bibfnamefont{A.}~\bibnamefont{Sch{\"a}licke}},
  \bibinfo{author}{\bibfnamefont{S.}~\bibnamefont{Schumann}}, \bibnamefont{and}
  \bibinfo{author}{\bibfnamefont{G.}~\bibnamefont{Soff}},
  \bibinfo{journal}{Phys. Rev.} \textbf{\bibinfo{volume}{D70}},
  \bibinfo{pages}{114009} (\bibinfo{year}{2004}), \eprint{hep-ph/0409106}.

\bibitem[{\citenamefont{Krauss et~al.}(2005)\citenamefont{Krauss,
  Sch{\"a}licke, Schumann, and Soff}}]{Krauss:2005nu}
\bibinfo{author}{\bibfnamefont{F.}~\bibnamefont{Krauss}},
  \bibinfo{author}{\bibfnamefont{A.}~\bibnamefont{Sch{\"a}licke}},
  \bibinfo{author}{\bibfnamefont{S.}~\bibnamefont{Schumann}}, \bibnamefont{and}
  \bibinfo{author}{\bibfnamefont{G.}~\bibnamefont{Soff}},
  \bibinfo{journal}{Phys. Rev.} \textbf{\bibinfo{volume}{D72}},
  \bibinfo{pages}{054017} (\bibinfo{year}{2005}), \eprint{hep-ph/0503280}.

\end{thebibliography}

\end{document}